\pdfoutput=1
\documentclass{new_tlp}

\usepackage[utf8]{inputenc}
\usepackage{xcolor}
\usepackage{listings}
\usepackage{booktabs}
\usepackage{enumitem}
\usepackage{amsmath}
\usepackage{mathtools}
\usepackage{tikz}
\usetikzlibrary{arrows.meta,positioning}
\usepackage{amssymb}
\usepackage{hyperref}
\hypersetup{colorlinks=true, linkcolor=black!70!blue, urlcolor=blue!60!black,
  citecolor=black, pdftitle={Programming in FastLAS}, pdfauthor={}}

\definecolor{cmtgray}{gray}{0.45}
\lstdefinelanguage{fastlas}{
  morekeywords={not},
  keywordsprefix={\#},
  morecomment=[l]{\%},
  morestring=[b]",
  morestring=[b]',
  sensitive=true,
}
\lstdefinestyle{fastlas}{
  language=fastlas,
  keywordstyle=\mdseries, keywordstyle=[2]\bfseries,
  commentstyle=\color{cmtgray},
  stringstyle=\itshape,
  xleftmargin=2.6em,
}
\lstdefinestyle{result}{
  language={},
  xleftmargin=2.6em,
}
\lstdefinestyle{shell}{
  language={},
  xleftmargin=2.6em,
}
\makeatletter
\renewcommand\paragraph{\@startsection{paragraph}{4}{\z@}%
  {\medskipamount}{-0.6em}{\normalfont\normalsize\bfseries}}
\makeatother

\usepackage{xparse}
\NewDocumentCommand{\code}{v}{\texttt{#1}}

\usepackage{framed}
\newenvironment{tlpnote}[2]
  {\par\addvspace{\medskipamount}%
   \def\FrameCommand{{\vrule width #1}\hspace{10pt}}%
   \MakeFramed{\advance\hsize-\width\FrameRestore}%
   \noindent\textsc{#2.}\ \ignorespaces}
  {\endMakeFramed\par\addvspace{\medskipamount}}
\newenvironment{ilaspbox}{\begin{tlpnote}{0.4pt}{Difference from ILASP}}{\end{tlpnote}}
\newenvironment{oplbox}{\begin{tlpnote}{0.4pt}{OPL vs.\ NOPL}}{\end{tlpnote}}
\newenvironment{gotcha}{\begin{tlpnote}{2pt}{Pitfall}}{\end{tlpnote}}

\newcounter{exmp}
\newcommand{\exhead}[1]{%
  \par\addvspace{\medskipamount}\refstepcounter{exmp}%
  \noindent\textbf{Example \theexmp.}\ \emph{#1}%
  \addcontentsline{toc}{subsubsection}{Example \theexmp. #1}%
  \par\nobreak\smallskip}

\title[An Unofficial FastLAS Tutorial]{An Unofficial FastLAS Tutorial:\\ A Programmer's Guide}

\author[F. A. D'Asaro]
       {FABIO AURELIO D'ASARO\\
        \textit{University of Verona, Italy}\\
        \textit{University College London, United Kingdom}\\
        \email{fabioaurelio.dasaro@univr.it; fabio.d'asaro.14@ucl.ac.uk}}

\jdate{July 2026}
\pubyear{2026}
\pagerange{\pageref{firstpage}--\pageref{lastpage}}
\doi{S1471068400000000}

\usepackage{etoolbox}
\makeatletter
\def\@underjournal{}
\def\@j@urnal{}
\patchcmd{\@maketitle}
  {{{\affilsize\it submitted \@submitted; revised \@revised; accepted \@accepted}}\par}
  {}{}{\PackageWarning{fastlas}{could not remove the submitted/revised/accepted line}}
\makeatother

\begin{document}

\label{firstpage}
\maketitle

\begin{abstract}
\noindent FastLAS is a scalable system for Inductive Logic Programming (ILP): you give it
some background knowledge, a \emph{language bias}, and a set of \emph{examples}, and it
searches for a set of logic-program rules (a \emph{hypothesis}) that explains the examples.
These notes are a hands-on introduction to \emph{writing} FastLAS programs. They are organised
as a programmer's guide: syntax first, then a ladder of worked, numbered examples of
increasing difficulty. Every FastLAS example here has been run against FastLAS
\texttt{2.2.0} and shows the tool's actual output; the few Clingo and ILASP snippets are marked
as such. We keep theory to the minimum needed to write
correct programs; throughout, set-off notes flag where FastLAS differs from its sibling system
ILASP, and where the two learning algorithms (\texttt{-{}-opl} and \texttt{-{}-nopl}) behave
differently. The document is intended as an unofficial tutorial to FastLAS~\texttt{2.2.0}, not as
an official language specification.
\end{abstract}

\noindent\textit{Disclosure.} This is an unofficial guide written from a user, for the users.
The author is not affiliated with ILASP LTD and has never participated in the development of ILASP or
FastLAS.\par\medskip


\tableofcontents
\bigskip

\noindent\textit{How this guide is organised.} Section~\ref{sec:glance} gives a first task at a glance. Sections~\ref{sec:anatomy} to~\ref{sec:effective} are the tutorial proper: the anatomy of a \texttt{.las} file, the choice of learning algorithm, and how to write programs that stay fast and stay within the system's limits. Section~\ref{sec:classics} revisits the classic ASP
exercises as learning tasks, and Section~\ref{sec:practice} works two published applications, CAVIAR event
recognition and access-control policy learning, as case studies. The remaining sections are compact reference material to consult rather than read straight through, followed by exercises whose solutions are in \ref{app:solutions}.\par\bigskip


\section{Introduction}

\subsection{What FastLAS does}
FastLAS~\cite{fastlas} learns \emph{Answer Set Programs} (ASP)~\cite{clingo} from examples:
ordinary logic-programming rules of the form \code{head :- body.} You describe:
\begin{itemize}[nosep,leftmargin=1.4em]
  \item \textbf{Background knowledge} $B$: facts and rules that are always true;
  \item \textbf{A language bias} $M$ (\emph{mode declarations}): the vocabulary the learned rules may use;
  \item \textbf{Examples} $E$: snapshots of what should (and should not) be entailed;
  \item \textbf{A scoring function} (\emph{bias}): what makes one hypothesis ``better'' than another.
\end{itemize}
FastLAS returns the lowest-scoring hypothesis $H$ such that $B \cup H$ explains every example.
Two things characterise it: it is built to \emph{scale} to tasks with tens of thousands of
examples, and its notion of a \emph{best} hypothesis is yours to define rather than fixed in
advance.

\subsection{FastLAS, ILASP, and this guide}
FastLAS is built by Mark Law and collaborators at Imperial College London, on the same
\emph{Learning from Answer Sets} (LAS) foundations~\cite{ilasp,lawthesis} as
\href{https://www.ilasp.com/}{ILASP}~\cite{ilaspsystem}.
The two systems share almost all of their \emph{input} syntax (mode declarations, the
example format, typed variables), so if you know one, most of the other is familiar. FastLAS
trades generality for speed and adds custom scoring; ILASP is more general (recursion, choice
rules, preferences). We point out the concrete differences in \textsc{Difference from ILASP}
notes as we go. At the time of writing there is no other dedicated FastLAS syntax tutorial;
the closest general reference is the ILASP manual. This document is an unofficial,
example-driven guide to FastLAS~\texttt{2.2.0}, not a formal language specification. For the
official materials alongside it, see the FastLAS project page
\url{https://spike-imperial.github.io/FastLAS/}, the FastLAS installation and running notes
\url{https://spike-imperial.github.io/FastLAS/installation.html}, and the ILASP manual
\url{https://doc.ilasp.com/}.

\subsection{What you need}
\label{sec:whatyouneed}
Two executables, both of which must be on your \texttt{PATH}. FastLAS is distributed from its
project page, \url{https://spike-imperial.github.io/FastLAS/}, which carries the releases and
build instructions; this guide uses version \texttt{2.2.0}. FastLAS runs the
\href{https://potassco.org/clingo/}{Clingo} ASP solver as an external process, so Clingo must be
installed as well. Check both before starting, since a task cannot run without either:
\begin{lstlisting}[style=shell]
$ FastLAS --version
$ clingo --version
\end{lstlisting} A FastLAS task is usually a single text file, conventionally with the extension \texttt{.las},
though several files given on one command line are simply concatenated, which lets you factor out
a shared background or bias. FastLAS ships \emph{two} learning algorithms, chosen by
a mandatory command-line flag:
\begin{itemize}[nosep,leftmargin=1.4em]
  \item \texttt{-{}-opl}: the original algorithm (\emph{Observational Predicate Learning}), which
        assumes the predicate you are learning is one the examples actually talk about;
  \item \texttt{-{}-nopl}: the later algorithm (\emph{Non-Observational Predicate Learning}), which
        drops that assumption and so handles more tasks, at some cost in speed.
\end{itemize}
Section~\ref{sec:oplnopl} is devoted to the difference. For now, just know that \emph{every}
run needs one of these flags.

\noindent Keep the official references close by while reading: the project page above, the
FastLAS installation/running notes
\url{https://spike-imperial.github.io/FastLAS/installation.html}, and the ILASP manual
\url{https://doc.ilasp.com/}. When a command-line detail or corner-case matters, those are the
authoritative sources.

\begin{gotcha}
Running \texttt{FastLAS task.las} with \textbf{no} algorithm flag prints
\texttt{ERROR: usage: FastLAS [ -{}-opl | -{}-nopl ] file\_name} and does nothing. Always pass
\texttt{-{}-opl} or \texttt{-{}-nopl}.
\end{gotcha}

\subsection{Notation used in this manual}
\label{sec:notation}
Three kinds of monospaced block appear throughout. A \emph{task file}, or a fragment of one, is
printed on its own:
\begin{lstlisting}[style=fastlas]
#modeh(cycle).
#modeb(rain).
\end{lstlisting}
A \emph{command} is introduced by a dollar prompt, which you do not type. Everything on the line
after the prompt is what you enter at a shell:
\begin{lstlisting}[style=shell]
$ FastLAS --opl examples/ex01_cycle.las
\end{lstlisting}
The \emph{output} FastLAS writes in response is printed in the same way but without a prompt, so
a command and its result can be read as a transcript:
\begin{lstlisting}[style=shell]
cycle :- not rain.
\end{lstlisting}
In running text we set directives, atoms, rules and flags in typewriter font, as in
\code{#modeh}, \code{cycle :- not rain.} and \code{--opl}. Command-line options keep the two
hyphens FastLAS expects. Every task file shown here is named where it is first used, and all of them are in the
\texttt{examples/} directory of the repository accompanying this guide:
\begin{center}
\url{https://github.com/dasaro/fastlas_manual}
\end{center}
Paths such as \texttt{examples/ex01\_cycle.las} are relative to a clone of that repository.


\section{FastLAS at a glance}
\label{sec:glance}

\subsection{The shape of a learning task}
A \texttt{.las} file is a plain-text file mixing up to four kinds of content, in any order and
not all of them required (the first task below has no background knowledge at all):
\begin{center}
\begin{tabular}{@{}ll@{}}
\toprule
\textbf{Background knowledge} & ordinary ASP facts/rules (\code{p :- q.}) \\
\textbf{Mode declarations}    & \code{#modeh(...)}, \code{#modeb(...)}, \code{#maxv(N)} \\
\textbf{Examples}             & \code{#pos(...)}, \code{#neg(...)} \\
\textbf{Scoring bias}         & \code{#bias("...")} \\
\bottomrule
\end{tabular}
\end{center}
Comments start with \code{
\texttt{//} or \texttt{\#}).

The rest of this section says what each of the four means, and what exactly FastLAS computes from
them. It is the only place in the guide with any formal machinery, and there is little of it: four
definitions and one optimisation problem.

\subsection{The four ingredients}
\label{sec:ingredients}

\paragraph{Background knowledge $B$.}
What you already know, written as an ordinary ASP program: facts such as \code{bird(tweety).} and
rules such as \code{flies(X) :- bird(X), not penguin(X).} It is fixed, it is never learned, and it
is the same for every example. Formally it is a set of normal rules, read under the stable-model
(answer set) semantics of \citeN{clingo}: a program does not have \emph{a} meaning but a set of
\emph{answer sets}, each one a consistent set of atoms that the program justifies. A program with
no choices, like a set of facts and non-recursive rules, has exactly one.

\paragraph{Language bias $M$: the mode declarations.}
What the learned rules are \emph{allowed} to say. \code{#modeh(A)} declares that the atom \code{A}
may appear as a rule head; \code{#modeb(L)} declares that the literal \code{L} may appear in a rule
body. Inside them, \code{var(t)} stands for a variable of type \code{t} and \code{const(t)} for a
constant of that type, so one declaration stands for many concrete literals. \code{#maxv(N)} caps
the number of distinct variables in any one rule.

Together these generate the \emph{search space} (or hypothesis space) $S_M$: the set of every rule
\[
  h \leftarrow b_1, \dots, b_n
\]
such that $h$ is an instance of some \code{#modeh}, each $b_i$ is an instance of some
\code{#modeb}, and the rule uses at most \code{#maxv} distinct variables. This set is what FastLAS
searches. A \emph{hypothesis} $H$ is any subset of it, $H \subseteq S_M$; nothing outside $S_M$ can
ever be learned, which is why an over-tight mode bias produces \code{UNSATISFIABLE} and an
over-loose one produces a long wait.

It helps to see $S_M$ written out once. ILASP, FastLAS's sibling system, will do exactly that: its
\code{-s} flag prints the whole hypothesis space generated by a mode bias. This is an
\textbf{ILASP} facility, not a FastLAS one, and we borrow it here purely to make the definition
concrete. Take this mode bias, which is all that \code{examples/ex29_space_ilasp.las} contains:
\begin{lstlisting}[style=fastlas]
bird(a). bird(b).
#modeh(flies(var(bird))).
#modeb(winged(var(bird))).
#modeb(penguin(var(bird))).
#maxv(1).
\end{lstlisting}
One head mode, two body modes, one variable. ILASP enumerates what that generates:
\begin{lstlisting}[style=shell]
$ ILASP --version=4 -s examples/ex29_space_ilasp.las
1 ~  :- winged(V1).
1 ~  :- penguin(V1).
2 ~ flies(V1) :- penguin(V1).
2 ~  :- penguin(V1); not winged(V1).
2 ~ flies(V1) :- winged(V1).
2 ~  :- winged(V1); penguin(V1).
2 ~  :- winged(V1); not penguin(V1).
3 ~ flies(V1) :- penguin(V1); not winged(V1).
3 ~ flies(V1) :- winged(V1); not penguin(V1).
3 ~ flies(V1) :- winged(V1); penguin(V1).
\end{lstlisting}
Ten candidate rules, each tagged with its length. Every combination of the declared literals
appears, positive and negated, and so do the headless constraints that ILASP is also willing to
learn. That set is what ``search space'' means, and you can see how quickly it would grow with one
more body mode or one more permitted variable.

FastLAS goes about it differently, and this is the heart of why it scales. It never builds $S_M$.
Instead it works example by example: for each one it computes a small set of candidate rules that
is \emph{sufficient} in the sense that some optimal hypothesis can still be assembled from it, and
searches only that~\cite{fastlas}. The space is defined by your modes exactly as above, but it is
approached from the data rather than enumerated in advance. What \code{--space-size} reports is the
size of the candidate set FastLAS actually kept, so on the same modes with a single example it
prints
\begin{lstlisting}[style=shell]
$ FastLAS --opl --space-size examples/ex29_space_fastlas.las
% SPACE SIZE: 1
flies(V0) :- bird(V0).
\end{lstlisting}
one, not ten. The number therefore answers ``how much did FastLAS have to consider here'', not
``how large is $S_M$''. It is still the right number to watch when a task is slow, because it is
the one that grows when your bias is too loose (Section~\ref{sec:perf}).

\paragraph{Examples $E$.}
What should and should not follow. An example is a four-slot \emph{context-dependent partial
interpretation}:
\begin{lstlisting}[style=fastlas]
#pos( id , { inclusions } , { exclusions } , { context } ).
\end{lstlisting}
The \emph{context} is a small ASP program describing one scenario, and it is added to the
background for that example only. The \emph{inclusions} and \emph{exclusions} are sets of atoms
saying what must, and must not, hold in that scenario. Write $B \cup C \cup H$ for the background,
this example's context and the hypothesis taken together. Then:
\begin{itemize}[nosep,leftmargin=1.4em]
  \item a \textbf{positive} example is \emph{covered} when \textbf{some} answer set $A$ of
        $B \cup C \cup H$ satisfies $\mathit{inc} \subseteq A$ and $\mathit{exc} \cap A =
        \emptyset$;
  \item a \textbf{negative} example is \emph{covered} when \textbf{no} answer set does.
\end{itemize}
``Some'', not ``every'': coverage is \emph{brave}. If the background admits several answer sets, a
positive example is satisfied as soon as one of them fits, even if the others do not.

\exhead{Brave coverage}
\label{ex:brave}
The background below has two answer sets, $\{\code{x},\code{p}\}$ and $\{\code{y}\}$. Only the
first contains \code{q}, and that is enough for \code{e1}.
\begin{lstlisting}[style=fastlas]
% Positive examples are covered BRAVELY: an inclusion need hold in only
% SOME answer set, not all of them. The background has two answer sets,
% {x,p} and {y}; only the first contains q, and that is enough for e1
% to be covered.
1 { x ; y } 1.
p :- x.
#modeh(q).
#modeb(p).
#pos(e1, {q}, {}, {}).
% context forces y, so p fails and q must not hold
#pos(e2, {}, {q}, { :- x. }).
#bias("penalty(1, body(X)) :- in_body(X).").
\end{lstlisting}
\begin{lstlisting}[style=shell]
$ FastLAS --opl examples/ex25_brave.las
q :- p.
\end{lstlisting}

\noindent An example with no weight is \emph{hard}: it must be covered. Attaching a penalty with
\code{@} makes it \emph{noisy}, and FastLAS may then leave it uncovered and pay that penalty
instead, which is what lets it tolerate mislabelled data.

\paragraph{Scoring bias.}
Which hypothesis you want, when several cover the examples. A \code{#bias("...")} string is a small
ASP program that assigns a cost to a candidate rule by inspecting it through the reserved
predicates \code{in_head/1} and \code{in_body/1}. The familiar pair
\begin{lstlisting}[style=fastlas]
#bias("penalty(1, head)    :- in_head(X).").
#bias("penalty(1, body(X)) :- in_body(X).").
\end{lstlisting}
charges one point per literal, which makes ``best'' mean ``shortest''. Section~\ref{sec:bias}
shows costs that no fixed objective could express.

\begin{gotcha}
\textbf{Without a \texttt{\#bias} there is no objective at all.} Every rule then costs zero, every
covering hypothesis is equally optimal, and \code{--score-only} duly reports \code{0}. FastLAS
still returns \emph{something}, but nothing is minimising length on your behalf, and which of
several covering hypotheses you get is not something you should rely on. Every worked example in
this guide supplies a scoring bias explicitly.
\end{gotcha}

\subsection{What FastLAS computes}
\label{sec:whatitcomputes}
Putting those together: given $B$, $M$, $E$ and a scoring bias, FastLAS searches for a hypothesis
$H \subseteq S_M$ that
\begin{enumerate}[nosep,leftmargin=1.6em]
  \item covers every \emph{hard} example, and
  \item minimises
        \[
          \mathcal{S}(H) \;=\; \sum_{r \in H} \mathit{cost}(r)
                          \;+\; \smashoperator{\sum_{e \,\in\, \mathit{uncovered}}} \mathit{penalty}(e)
        \]
        where the first sum is the cost your \code{#bias} assigns to the rules you keep, and the
        second is the price of the noisy examples you chose not to cover.
\end{enumerate}
and prints it, one rule per line. If no $H \subseteq S_M$ covers the hard examples, it prints
\code{UNSATISFIABLE}. Both halves of that sum are observable: \code{--score-only} prints
$\mathcal{S}(H)$ for the returned hypothesis.

\exhead{The two halves of the score}
\label{ex:noisy-score}
Here \code{e1} and \code{e2} contradict each other. \code{e1} is hard, so it must be covered;
\code{e2} carries penalty~1, so FastLAS covers \code{e1}, leaves \code{e2} uncovered, and pays.
\begin{lstlisting}[style=fastlas]
% e1 and e2 contradict each other. e1 is hard, so it must be covered.
% e2 carries penalty 1, so FastLAS may leave it uncovered and pay 1.
#modeh(p).
#modeb(a).
#pos(e1,   {p}, {},  { a. }).
#pos(e2@1, {},  {p}, { a. }).
#bias("penalty(1, head)    :- in_head(X).").
#bias("penalty(1, body(X)) :- in_body(X).").
\end{lstlisting}
\begin{lstlisting}[style=shell]
$ FastLAS --opl examples/ex26_noisy.las
p.
$ FastLAS --opl --score-only examples/ex26_noisy.las
2
\end{lstlisting}
The \code{2} is one point for the learned rule \code{p.} and one for the abandoned example.

\subsection{Your first task}
Let us learn the rule ``I cycle to work unless it rains''. We will observe two days.

\exhead{Hello, FastLAS (propositional)}
\label{ex:cycle}
\begin{lstlisting}[style=fastlas]
#modeh(cycle).
#modeb(rain).
#modeb(not rain).

#pos(d1, {cycle}, {}, {}).
#pos(d2, {}, {cycle}, {rain.}).

#bias("penalty(1, head) :- in_head(X).").
#bias("penalty(1, body(X)) :- in_body(X).").
\end{lstlisting}

\noindent Read it top to bottom:
\begin{itemize}[nosep,leftmargin=1.4em]
  \item \code{#modeh(cycle).} says the head of a learned rule may be the atom \code{cycle}.
  \item \code{#modeb(rain).} and \code{#modeb(not rain).} let rule bodies use \code{rain}
        or its negation \code{not rain}.
  \item \code{#pos(d1, {cycle}, {}, {}).} is example \code{d1}: in the scenario with empty
        context \code{{}} (a clear day), \code{cycle} must hold. The four slots are
        \emph{id}, \emph{inclusions}, \emph{exclusions}, \emph{context}.
  \item \code{#pos(d2, {}, {cycle}, {rain.}).} is example \code{d2}: when the context
        contains the fact \code{rain}, \code{cycle} must \emph{not} hold (it is in the
        exclusions). Note the full stop in \code{{rain.}}: the fourth slot is an ASP
        \emph{program}, so every fact in it ends with a dot, whereas the two set slots hold bare
        atoms and do not. Omitting that dot is a syntax error, and it is the commonest slip when
        copying this line as a template.
  \item The two \code{#bias} lines say ``charge 1 point per head atom and 1 point per body
        atom'', i.e.\ prefer the shortest rule.
\end{itemize}

\noindent Run it:
\begin{lstlisting}[style=shell]
$ FastLAS --opl examples/ex01_cycle.las
\end{lstlisting}
\begin{lstlisting}[style=result]
cycle :- not rain.
\end{lstlisting}
FastLAS found exactly the intended rule. Because this task is fully observational, both
algorithms agree: \texttt{FastLAS -{}-nopl examples/ex01\_cycle.las} prints the same thing.

\subsection{Running FastLAS from the command line}
The basic invocation is
\begin{lstlisting}[style=shell]
$ FastLAS --opl  task.las    # original FastLAS algorithm
$ FastLAS --nopl task.las    # FastNonOPL (non-observational learning)
\end{lstlisting}
The remaining flags (\texttt{-{}-version}, \texttt{-{}-help}, \texttt{-{}-debug},
\texttt{-{}-force-safety}, \texttt{-{}-threads N}, \texttt{-{}-timeout T}) are summarised in
Section~\ref{sec:running}.
The learned hypothesis is printed to standard output, one rule per line. If no hypothesis
explains the examples, FastLAS prints \texttt{UNSATISFIABLE}.

\section{The anatomy of a \texttt{.las} file}
\label{sec:anatomy}
Every FastLAS task is built from the same pieces: \emph{background knowledge}, a \emph{mode-declaration} language bias, \emph{examples}, and a \emph{scoring} function. This section covers each in turn: the whole input language in one place.

\subsection{Background knowledge: ASP in one page}
\label{sec:bg}
Background knowledge is an ASP program in the subset accepted by FastLAS, close to Clingo for
ordinary facts and rules but not identical to full Clingo syntax. If you know Prolog/Datalog it
will look familiar. The pieces you need:

\begin{center}
\begin{tabular}{@{}ll@{}}
\toprule
\textbf{Construct} & \textbf{Example} \\
\midrule
Fact               & \code{bird(tweety).} \\
Rule               & \code{flies(X) :- bird(X), not penguin(X).} \\
Integrity constraint & \code{:- smaller, greater.} \quad(``never both'') \\
Choice rule        & \code{1 { a ; b } 1 :- h.} \\
Interval / range   & \code{num(1..100).} \\
Arithmetic         & \code{+ - * / \ **} \ and \ \code{|X|} \ (absolute value) \\
Comparisons        & \code{=  !=  <  <=  >  >=} \\
Comment            & \code{
\bottomrule
\end{tabular}
\end{center}

One word on \code{not}, since it is the source of most early confusion. It is \emph{negation as
failure}, not classical negation: \code{not penguin(X)} does not assert that \code{X} is provably
not a penguin, only that nothing in the program manages to derive that it is. Absence of evidence
is treated as evidence of absence, which is what makes these programs \emph{non-monotonic}: adding
the fact \code{penguin(tweety)} later can retract a conclusion that held before.

Variables start with an \textbf{uppercase} letter (\code{X}, \code{Time}); predicate and
constant names start \textbf{lowercase} (\code{bird}, \code{tweety}); \code{_} is an anonymous
variable; strings are double- or single-quoted. Ranges like \code{num(1..100)} and arithmetic
such as \code{X+1} are expanded by Clingo. You will typically put universal rules and type
definitions in the background, and the specifics of each example in that example's
\emph{context} (Section~\ref{sec:context}).

\paragraph{A program, built up.}
It is worth writing one small program before going further, because the background knowledge of
a learning task is exactly this and nothing more. Facts state what is true, and a range states
many facts at once, so \code{num(1..5).} is five facts. A rule derives new atoms from old, and
its variables are read as ``for every'':
\begin{lstlisting}[style=fastlas]
composite(N) :- N = 1..5, I = 2..N-1, N\I = 0.
\end{lstlisting}
This says that \code{N} is composite when some \code{I} strictly between \code{1} and \code{N}
divides it, \code{\} being the remainder operator. Negation as failure then gives the primes,
with \code{not composite(N)} holding for every \code{N} the first rule failed to derive:
\begin{lstlisting}[style=fastlas]
prime(N) :- N = 2..5, not composite(N).
\end{lstlisting}
\begin{lstlisting}[style=shell]
$ clingo examples/asp/asp01_primes.lp
prime(2) prime(3) prime(5)
\end{lstlisting}
That is the whole of what a FastLAS background program does: derive further atoms from the facts
of the task. The remaining ASP constructs, choice rules and constraints, do something different.
They describe a \emph{space} of possible answers rather than deriving one, which is how Clingo is
used to solve puzzles and how ILASP describes hypotheses. FastLAS does accept them in the
background, with one restriction: a choice rule must carry \emph{both} bounds, so
\code{1 { x ; y } 1.} parses but \code{{ x }.} is a syntax error. What FastLAS will not do is
\emph{learn} them; they can only be background. Section~\ref{sec:classics} shows what that means
in practice: the classic guess-and-test
programs stay in Clingo, and what FastLAS contributes is learning the rules inside them.

\begin{gotcha}
\textbf{FastLAS's ASP is not full Clingo.} Its parser rejects \emph{conditional literals}, the
\code{:} ``for every'' construct. A background rule like
\begin{lstlisting}[style=fastlas]
defended(X) :- arg(X), out(Y) : att(Y,X).      % rejected
\end{lstlisting}
fails with \code{syntax error, unexpected T_COLON}. Rewrite it with an auxiliary predicate and
negation. Name the \emph{opposite} property and derive yours from its absence:
\begin{lstlisting}[style=fastlas]
spoiled(X)  :- att(Y,X),
        not out(Y).            % some attacker of X is not out
defended(X) :- arg(X),
        not spoiled(X).        % no such attacker exists
\end{lstlisting}
This parses, and \code{defended} is then usable as an ordinary \code{#modeb} feature.
\end{gotcha}

\paragraph{Dialect limits: what FastLAS's ASP won't accept.}
Conditional literals (just above) are one instance of a general rule: FastLAS's own lexer knows
only a fixed set of \code{#}-keywords and constructs, so most Clingo ``extras'' are rejected
outright. There is no fall-through to Clingo for them. (To see what FastLAS \emph{does} produce,
\code{--output-solve-program} is described in Section~\ref{sec:running}.)
\begin{center}
\begin{tabular}{@{}ll@{}}
\toprule
You write & FastLAS says \\
\midrule
\code{#show sel/0.}                          & \code{Unknown token: '#'} \\
\code{q :- #count{...} > 0.} (aggregates)    & \code{Unknown token: '#'} \\
\code{#minimize{...}.} / \code{#maximize}    & \code{Unknown token: '#'} \\
\code{:~ sel. [1@1]} (weak constraint)       & \code{syntax error, unexpected T_COLON} \\
\code{p(X) :- q(Y) : r(X,Y).} (cond.\ lit.)  & \code{syntax error, unexpected T_COLON} \\
\bottomrule
\end{tabular}
\end{center}
Work within it: compute what you need with plain normal rules and negation (Section~\ref{sec:bg}),
and supply any optimisation (\code{#minimize}) or weak constraints at \emph{deploy} time around the
learned theory rather than inside the task. This is the generate-and-constrain pattern of
Section~\ref{sec:limits}.

\subsection{The language bias: mode declarations}
\label{sec:modes}
Mode declarations define the \emph{hypothesis space}: the set of rules FastLAS is allowed to
build. A learned rule takes its head from a \code{#modeh} declaration and each body literal
from a \code{#modeb} declaration.

\paragraph{\texttt{\#modeh} and \texttt{\#modeb}.}
In the propositional case (Example~\ref{ex:cycle}) the declarations are just atoms. The power
comes from \emph{typed variables}.

\paragraph{Typed variables: \texttt{var(t)}.}
Inside a mode, \code{var(t)} is a placeholder for a variable of ``type'' \code{t}. The type
\code{t} is an ordinary unary background predicate \code{t/1} that enumerates the type's values.
For instance \code{#modeh(flies(var(animal)))} says ``a learned rule may have head
\code{flies(X)} where \code{X} ranges over things satisfying \code{animal(X)}''.

\begin{gotcha}
\textbf{Safety.} As in ASP, every variable in a rule must be \emph{bound by a positive body
literal}. A rule like \code{flies(X) :- not penguin(X).} is unsafe on its own, because \code{X}
occurs only under negation. FastLAS does not reject such a rule: it appends the type atom
(e.g.\ \code{animal(X)}) to the body, which makes it safe, and that is why you see the type atom
in every learned rule. Example~\ref{ex:flies-exc} learns exactly such a rule,
\code{flies(V0) :- not flightless(V0), animal(V0).}, whose only declared body literal is negated.
If you would rather FastLAS satisfied safety from your \emph{declared} modes instead of leaning
on the type atom, pass \code{--force-safety}; on that same example it returns the longer
\code{flies(V0) :- winged(V0), not flightless(V0), animal(V0).} The flag changes the answer, so
do not reach for it casually while debugging.
\end{gotcha}

\exhead{Typed variables and generalisation}
\label{ex:flies-gen}
Two birds, both winged, both fly. What is the simplest rule?
\begin{lstlisting}[style=fastlas]
animal(eagle).  animal(sparrow).

#modeh(flies(var(animal))).
#modeb(winged(var(animal))).
#maxv(1).
#bias("penalty(1, X) :- in_head(X).").
#bias("penalty(1, X) :- in_body(X).").

#pos(p1, {flies(eagle)},   {}, { winged(eagle).   }).
#pos(p2, {flies(sparrow)}, {}, { winged(sparrow). }).
\end{lstlisting}
\begin{lstlisting}[style=shell]
$ FastLAS --opl examples/ex02_flies_general.las
\end{lstlisting}
\begin{lstlisting}[style=result]
flies(V0) :- animal(V0).
\end{lstlisting}
Note two things. First, FastLAS names the learned variable \code{V0}. Second, with no
counter-examples the \emph{most general} rule wins: ``all animals fly''. The body mode
\code{winged} was available but not needed, and the shortest hypothesis is preferred. To learn
something less trivial we need examples that push back, and the per-example data must
live in the \emph{context} (note \code{winged(eagle)} sits in \code{p1}'s context). This is the
single most important idiom in FastLAS; we return to it in Section~\ref{sec:context}.

\paragraph{Constants: \texttt{const(t)}.}
\code{const(t)} is a placeholder for a specific \emph{constant} of type \code{t} to be baked
into the rule (rather than a variable). In FastLAS the candidate constants are drawn from
\code{t(C)} facts, whether those facts sit in the background or in a positive example's
context.

\exhead{A learned constant with \texttt{const(t)}}
\label{ex:const}
``\code{sel} holds exactly when the chosen digit is 2.''
\begin{lstlisting}[style=fastlas]
#modeh(sel).
#modeb(chosen(const(digit))).
#bias("penalty(1, X) :- in_head(X).").
#bias("penalty(1, X) :- in_body(X).").

#pos(p1, {sel}, {},    { chosen(2). digit(2). }).
#pos(p2, {},    {sel}, { chosen(1). digit(1). }).
#pos(p3, {},    {sel}, { chosen(3). digit(3). }).
\end{lstlisting}
\begin{lstlisting}[style=shell]
$ FastLAS --opl examples/ex05_const_select.las
\end{lstlisting}
\begin{lstlisting}[style=result]
sel :- chosen(2).
\end{lstlisting}
The constant \code{2} was selected because \code{digit(2)} appears in a positive example's
context. The other examples rule out \code{chosen(1)} and \code{chosen(3)}.

\begin{ilaspbox}
In ILASP, \code{const(t)} values come from \code{#constant(t, v).} FastLAS has no
\code{#constant} directive. Instead, supply the constants as ordinary \code{t(v)} facts in
your positive example contexts (or background).
\end{ilaspbox}

\paragraph{Negation as failure in modes.}
To allow a learned rule to use \code{not p}, declare it explicitly:
\code{#modeb(not penguin(var(animal)))}. A body mode with no \code{not} counterpart can never
appear negated.

\begin{ilaspbox}
ILASP \emph{auto-generates} the negated version of each \code{#modeb} literal, and you write the
option \code{(positive)} to switch that off. FastLAS is the other way round: negation is
opt-\emph{in}, declared with a separate \code{#modeb(not ...)}. FastLAS also does not support the
ILASP mode-option tuples \code{(positive)}, \code{(anti_reflexive)}, \code{(symmetric)}.
\end{ilaspbox}

\paragraph{Recall and \texttt{\#maxv}.}
A leading integer in a body mode is a \emph{recall} bound, written
\code{#modeb(1, actual(var(number), var(number)))}, which in other systems caps how many times that
literal may appear in one rule. In FastLAS the control that matters is \code{#maxv(N)}, which caps
the number of distinct variables in any single learned rule: it is the main lever you have on the
size of the search, and Section~\ref{sec:perf} measures its effect.

\paragraph{Comparisons and arithmetic as body modes.}
Relational operators can themselves be body modes, letting learned rules contain inequalities:
\begin{lstlisting}[style=fastlas]
#modeb(var(number) < var(number)).
#modeb(var(number) >= var(number)).
\end{lstlisting}
Arithmetic in the \emph{contexts} of examples is evaluated by Clingo before learning starts, and
Example~\ref{ex:arith} relies on that. Declaring arithmetic as a body \emph{mode} is a different
matter: FastLAS is not built to learn arithmetic relations. Write the numeric argument as
\code{const(t)} rather than as a literal, \code{#modeb(sz(var(thing), const(t)))} rather than
\code{#modeb(sz(var(thing), 3))}, and reach for \code{num_var} (below) when what you want is a
numeric \emph{bound}.

\paragraph{Learning numeric thresholds: \texttt{num\_var}.}
\code{var(t)} and \code{const(t)} range over \emph{symbolic} values. For \emph{numeric} data
FastLAS has a third placeholder, \code{num_var(t)}, that does something the others cannot: it makes
FastLAS \emph{synthesise the comparison bounds itself}. You never write the threshold. FastLAS
searches for the \code{>=}/\code{<=} constants that fit the examples. As in the examples below,
the task should define the relevant numeric domain and provide the observed values in the example
contexts.

\exhead{Synthesising a numeric interval}
\label{ex:numvar}
Four cars: two accepted (speeds 60 and 80) and two rejected (30 and 110), so the accepted band is
bounded on \emph{both} sides.
\begin{lstlisting}[style=fastlas]
car(c1). car(c2). car(c3). car(c4).
speed_reading(0..120).
#modeh(ok(var(car))).
#modeb(observed(var(car), num_var(speed_reading))).
#maxv(1).
#pos(p1, {ok(c1)}, {},       { observed(c1, 60). }).
#pos(p2, {ok(c2)}, {},       { observed(c2, 80). }).
#pos(n1, {},       {ok(c3)}, { observed(c3, 30). }).
#pos(n2, {},       {ok(c4)}, { observed(c4, 110). }).
#bias("penalty(1, body(X)) :- in_body(X).").
\end{lstlisting}
\begin{lstlisting}[style=shell]
$ FastLAS --nopl examples/ex17_numvar.las
\end{lstlisting}
\begin{lstlisting}[style=result]
ok(V0) :- observed(V0,V_0_speed_reading), V_0_speed_reading >= 60,
        V_0_speed_reading <= 80, car(V0).
\end{lstlisting}
FastLAS invented \emph{both} bounds (\code{>= 60} and \code{<= 80}) from the data; no
comparison was declared as a mode. The synthesised numeric variable is named \code{V_<i>_<type>}.
With negatives on only one side you would get a single bound (e.g.\ \code{V >= 70}); a plain
\code{var} in place of \code{num_var} synthesises nothing and the task is \code{UNSATISFIABLE}.

\paragraph{Several numeric conditions at once.}
Two flags govern how many numeric comparisons a rule may carry:
\begin{itemize}[nosep,leftmargin=1.4em]
  \item \code{--max-conditions N} (default \textbf{1}): the number of distinct bounded numeric
        variables allowed \emph{per rule}. One variable's \code{>=} and \code{<=} together count as one.
  \item \code{--num-var-count N} (default 1): the number of numeric slots available \emph{per
        numeric type}.
\end{itemize}
To bound two \emph{different} quantities you must lift the default cap:
\begin{lstlisting}[style=fastlas]
car(c1). car(c2). car(c3).
speed_reading(0..120).  weight_reading(0..3000).
#modeh(ok(var(car))).
#modeb(fast(var(car),  num_var(speed_reading))).
#modeb(heavy(var(car), num_var(weight_reading))).
#maxv(1).
#pos(p1, {ok(c1)}, {},       { fast(c1,80). heavy(c1,1500). }).
#pos(n1, {},       {ok(c2)}, { fast(c2,40). heavy(c2,1500). }).
#pos(n2, {},       {ok(c3)}, { fast(c3,80). heavy(c3,500).  }).
#bias("penalty(1, body(X)) :- in_body(X).").
\end{lstlisting}
\begin{lstlisting}[style=shell]
% the default cap of 1 makes this UNSATISFIABLE
$ FastLAS --nopl examples/ex17_numvar_multi.las
$ FastLAS --nopl --max-conditions 2 examples/ex17_numvar_multi.las
\end{lstlisting}
\begin{lstlisting}[style=result]
ok(V0) :- fast(V0,V_0_speed_reading), heavy(V0,V_0_weight_reading),
        V_0_weight_reading >= 1500, V_0_speed_reading >= 80, car(V0).
\end{lstlisting}
At the default cap of one numeric condition the task is \code{UNSATISFIABLE};
\code{--max-conditions 2} lets the rule bound both speed and weight.

\subsection{Examples: context-dependent partial interpretations}
\label{sec:context}
An example is a \emph{Context-Dependent Partial Interpretation} (CDPI). Its four slots are:
\begin{center}
\begin{lstlisting}[style=fastlas]
#pos( id , { inclusions } , { exclusions } , { context } ).
%              bare atoms        bare atoms       an ASP program:
%            (no dots)       (no dots)      every fact ends in a dot
\end{lstlisting}
\end{center}
Meaning: \emph{there is an answer set of $B \cup H$ together with the \textbf{context}, that
contains every \textbf{inclusion} atom and none of the \textbf{exclusion} atoms.} The
\emph{context} is a little ASP program local to that example: the scenario the example is
about. The \emph{id} may carry a penalty (\emph{Weighted (noisy) examples}, below); it can even be
omitted.

The context slot is the one that does most of the work in real tasks, so it is worth dwelling on.
Everything you put in it is added to the background \emph{for that example only}, which means one
file can hold thousands of unrelated situations without them interfering. A medical task might give
each patient their own context; an access-control task gives each request one
(Example~\ref{ex:policy}); the CAVIAR study of Section~\ref{sec:caviar} gives each video frame
one. Contexts may
contain rules, not just facts, so a scenario can carry its own local definitions:
\begin{lstlisting}[style=fastlas]
% two patients, two scenarios, one file
#pos(p1, {treat(ann)}, {},
     { fever(ann). cough(ann). ill(X) :- fever(X), cough(X). }).
#pos(p2, {}, {treat(bob)},
     { cough(bob). ill(X) :- fever(X), cough(X). }).
\end{lstlisting}
Two things follow. Anything shared by \emph{every} example belongs in the background instead, where
it is written once; and anything that varies between scenarios must be in the context, because an
example is otherwise indistinguishable from any other with the same inclusions and exclusions.

\begin{gotcha}
\textbf{Put the scenario in the context.} The most common beginner mistake is to leave contexts
empty (\code{{}}) and pile all the data into global background. Examples grounded only through
global facts typically come back \texttt{UNSATISFIABLE}. Rule of thumb: universal rules and type
definitions go in the background; the facts describing \emph{this} example go in \emph{its}
context.
\end{gotcha}

\paragraph{Exclusions: expressing the negative case.}
Putting an atom in the \emph{exclusions} of a positive example says ``in this scenario, that
atom must \emph{not} be derivable''. This is the usual way to express a counter-example.

\exhead{Rules with exceptions (inclusions vs.\ exclusions)}
\label{ex:flies-exc}
``Winged animals fly, unless they are flightless.''
\begin{lstlisting}[style=fastlas]
animal(eagle).  animal(ostrich).  animal(sparrow).

#modeh(flies(var(animal))).
#modeb(winged(var(animal))).
#modeb(not flightless(var(animal))).
#maxv(1).
#bias("penalty(1, X) :- in_head(X).").
#bias("penalty(1, X) :- in_body(X).").

#pos(p1, {flies(eagle)},   {},            { winged(eagle).   }).
#pos(p2, {flies(sparrow)}, {},            { winged(sparrow). }).
#pos(p3, {}, {flies(ostrich)},
     { winged(ostrich). flightless(ostrich). }).
\end{lstlisting}
\begin{lstlisting}[style=shell]
$ FastLAS --opl examples/ex03_flies_exception.las
\end{lstlisting}
\begin{lstlisting}[style=result]
flies(V0) :- not flightless(V0), animal(V0).
\end{lstlisting}
Example \code{p3} is a positive example whose \emph{exclusion} \code{{flies(ostrich)}} forbids
the ostrich from flying; its context supplies \code{flightless(ostrich)}. That single
counter-example forces the \code{not flightless(V0)} literal into the rule.

\paragraph{Negative examples \texttt{\#neg}.}
A negative example states that \emph{no} answer set of $B \cup H$ with the given context
satisfies the inclusion/exclusion pattern. Syntactically it looks like \code{#pos} but with
\code{#neg}. Where a positive example says \emph{this must be possible}, a negative one says
\emph{this must be impossible}, and that is a much stronger demand: one stray answer set is enough
to violate it. Negative examples are how you stop a learner settling for a rule that is too
permissive.

\exhead{A negative example rules out the over-general rule}
\label{ex:neg-penguin}
Polly flies and Tweety, a penguin, must not. The positive example alone is satisfied by
\code{flies(V0) :- bird(V0).}, since nothing yet forbids Tweety from flying; the negative example
is what forces the exception into the rule.
\begin{lstlisting}[style=fastlas]
% A negative example says: no answer set may look like this. Here it is
% what stops the learner settling for "all birds fly".
bird(tweety). bird(polly). penguin(tweety).

#modeh(flies(var(bird))).
#modeb(bird(var(bird))).
#modeb(not penguin(var(bird))).
#maxv(1).

#pos(p1, {flies(polly)},  {}, {}).   % polly does fly
#neg(n1, {flies(tweety)}, {}, {}).   % tweety must NOT fly

#bias("penalty(1, head)    :- in_head(X).").
#bias("penalty(1, body(X)) :- in_body(X).").
\end{lstlisting}
\begin{lstlisting}[style=shell]
$ FastLAS --nopl examples/ex27_neg_penguin.las
flies(V0) :- not penguin(V0), bird(V0).
\end{lstlisting}
Delete the \code{#neg} line and re-run: the answer degrades to \code{flies(V0) :- bird(V0).},
which covers the one positive example just as well and happens to be shorter. The negative example
is carrying the whole distinction.

\begin{gotcha}
\textbf{Run tasks containing \texttt{\#neg} with \texttt{-{}-nopl}.} Under \code{--opl} a task with
a negative example in it returns \code{UNSATISFIABLE}; the same file solves under \code{--nopl}:
\begin{lstlisting}[style=shell]
$ FastLAS --opl  examples/ex23_neg.las
UNSATISFIABLE
$ FastLAS --nopl examples/ex23_neg.las
cycle :- not rain.
\end{lstlisting}
Since most of this guide runs under \code{--opl}, the safe habit is to state counter-examples as
\emph{exclusions} of a \code{#pos} wherever you can, as Example~\ref{ex:flies-exc} does. Reach
for \code{#neg} only when you genuinely need ``no answer set may look like this'', and then pass
\code{--nopl}.

Two smaller points about the syntax. \code{#neg} \textbf{must} use the four-slot form: FastLAS
accepts \code{#pos} with either three slots (\emph{id}, inclusions, exclusions) or four, but
\code{#neg(e, {a}, {b}).} is a syntax error, and you want \code{#neg(e, {a}, {b}, {}).} with an
explicit, possibly empty, context. And the id may carry a penalty just as a positive example may,
so \code{#neg(e@5, ...)} is a noisy negative example.
\end{gotcha}

\paragraph{Computed inclusions and exclusions.}
The two set slots can also be filled \emph{by the background program} rather than written out.
\code{inclusion/1} and \code{exclusion/1} are reserved: whatever the background derives for them,
in the presence of an example's context, is added to that example's inclusion and exclusion sets.
So this task learns \code{flies} even though both slots of both examples are empty:
\begin{lstlisting}[style=fastlas]
% Computed inclusions and exclusions: the two set slots of every example
% are empty, and the background derives them instead (Section 3.3).
animal(a). animal(b).
inclusion(flies(X)) :- animal(X), goal(X).
exclusion(flies(X)) :- animal(X), not goal(X).
#modeh(flies(var(animal))).
#modeb(winged(var(animal))).
#pos(e1, {}, {}, { goal(a). winged(a). }).
#pos(e2, {}, {}, { }).
#bias("penalty(1, body(X)) :- in_body(X).").
\end{lstlisting}
\begin{lstlisting}[style=shell]
$ FastLAS --opl examples/ex24_computed_slots.las
flies(V0) :- winged(V0), animal(V0).
\end{lstlisting}
Rename the two predicates to anything else and the same task returns the empty hypothesis, which
is the quickest way to convince yourself the names are special. This is how tasks with thousands
of examples are written in practice: each example carries only its scenario, and two background
rules say what should follow. The CAVIAR study of Section~\ref{sec:caviar} is built entirely on
this mechanism.

\paragraph{Weighted (noisy) examples.}
Attach a penalty to an example id with \code{@}: \code{#pos(a1@1, {smaller}, {greater}, {...})},
and likewise \code{#neg(n1@1, ...)}. An unweighted example is \emph{hard} and must be covered; a
weighted one may be left uncovered at a cost equal to its penalty. Real data disagrees with itself,
and this is what lets FastLAS shrug off the disagreement instead of returning
\code{UNSATISFIABLE}: an outlier that would force a contorted rule can simply be bought off, if
buying it off is cheaper than the extra literals.

Because coverage and hypothesis cost are added together (Section~\ref{sec:whatitcomputes}), the
penalties are not merely labels. They decide the answer.

\exhead{The weights decide}
\label{ex:weights}
Two examples flatly contradict each other, and both are noisy. FastLAS keeps whichever is more
expensive to abandon.
\begin{lstlisting}[style=fastlas]
% Two examples that contradict each other, both noisy. FastLAS keeps the
% one that is more expensive to abandon: here e1, at the price of
% leaving e2 uncovered.
% Swap the two weights and the answer flips to the empty hypothesis.
#modeh(p).
#modeb(a).

#pos(e1@3, {p}, {},  { a. }).   % costs 3 to ignore
#pos(e2@1, {},  {p}, { a. }).   % costs 1 to ignore

#bias("penalty(1, head)    :- in_head(X).").
#bias("penalty(1, body(X)) :- in_body(X).").
\end{lstlisting}
\begin{lstlisting}[style=shell]
$ FastLAS --opl examples/ex28_weights.las
p.
$ FastLAS --opl --score-only examples/ex28_weights.las
2
\end{lstlisting}
It covered \code{e1} and paid \code{1} for abandoning \code{e2}: one point for the rule
\code{p.} and one for the abandoned example. Now swap the two weights, so that \code{e1} costs~1
to ignore and \code{e2} costs~3, and run it again. FastLAS returns \emph{nothing at all} with a
score of~1: the empty hypothesis costs nothing, covers \code{e2}, and abandons \code{e1} for a
single point. Same data, same modes, opposite conclusions.

\begin{ilaspbox}
The example format (the 4-tuple CDPI and the \code{id@penalty} weighting) is
byte-for-byte the same as ILASP, and so is the acceptance condition for a positive example:
\emph{brave}, meaning some answer set extends it. What differs is what each system asks of the
\emph{background}. \code{--opl} assumes $B \cup \textit{context}$ is categorical, one answer set
per example, which is what lets it be fast; ILASP makes no such assumption, and neither does
\code{--nopl}. See Section~\ref{sec:oplnopl}.
\end{ilaspbox}

\paragraph{A harder example: learning arithmetic.}
This example brings together typed numeric variables, arithmetic in contexts, and a
length bias.

\exhead{Learning \texttt{result(X)} from expressions}
\label{ex:arith}
\begin{lstlisting}[style=fastlas]
num(1..100).
#modeh(result(var(num))).
#modeb(expr(var(num))).
#maxv(1).
#bias("penalty(1, X) :- in_head(X).").
#bias("penalty(1, X) :- in_body(X).").


#pos(eg1, {result(5)}, { result(1) }, { expr(4+1). }).
#pos(eg2, {result(5)}, { }, { expr((4+1)). }).
#pos(eg3, {result(10)}, { }, { expr((4+1)*2). }).
#pos(eg4, {result(2)}, { }, { expr((4+1)/2). }).
#pos(eg5, {result(25)}, { }, { expr((4+1)**2). }).
#pos(eg6, {result(6)}, { }, { expr(|(4-7)*2|). }).


#pos(eg7, {result(6)}, { }, { expr(4+1*2). }).
#pos(eg8, {result(8)}, { }, { expr(4/1*2). }).
#pos(eg9, {result(1)}, { }, { expr(4-2-1). }).
#pos(eg10, {result(3)}, { }, { expr(4-(2-1)). }).
\end{lstlisting}
\begin{lstlisting}[style=shell]
$ FastLAS --opl examples/ex04_arithmetic.las
\end{lstlisting}
\begin{lstlisting}[style=result]
result(V0) :- expr(V0), num(V0).
\end{lstlisting}
Each example puts an arithmetic expression in its context (\code{expr(4+1)}, \code{expr((4+1)*2)},
\code{expr(|(4-7)*2|)}, \dots); Clingo evaluates the arithmetic, and FastLAS learns the single
rule that maps every expression to its value. Note again the auto-appended type atom
\code{num(V0)} that keeps the rule safe.

\subsection{Scoring: the \texttt{\#bias} mini-language}
\label{sec:bias}
This is FastLAS's signature feature. Each \code{#bias("...")} contains a fragment of an ASP
program that FastLAS evaluates \emph{per candidate rule} to compute that rule's \emph{cost}. The
total cost of a hypothesis is the sum of its rule costs (plus penalties for any uncovered
weighted examples); FastLAS returns a \emph{minimum-cost} hypothesis. The reserved vocabulary:
\begin{itemize}[nosep,leftmargin=1.4em]
  \item \code{penalty(W, Id)}: charge weight \code{W} (an integer, possibly negative, or an
        arithmetic expression) under identifier \code{Id}. Equal-\code{Id} penalties are counted
        once.
  \item \code{in_head(X)}: true iff atom \code{X} is in the candidate rule's head.
  \item \code{in_body(X)}: true iff literal \code{X} is in the candidate rule's body.
  \item plus arbitrary ASP: your own predicates, \code{#count}, arithmetic, comparisons\dots
This is not a contradiction of the dialect limits of Section~\ref{sec:bg}: the string inside
\code{#bias} is handed to Clingo, so Clingo's aggregates are available there even though the same
constructs are rejected in the background program.
\end{itemize}
The idiomatic ``prefer shortest rule'' scoring is
\begin{lstlisting}[style=fastlas]
#bias("penalty(1, head)      :- in_head(X).").
#bias("penalty(1, body(X))   :- in_body(X).").
\end{lstlisting}
Here \code{head} and \code{body(X)} are just penalty \emph{identifiers}; the special predicates
are \code{in_head}/\code{in_body}. This pair is the one you will write most often, and it is what
makes ``best'' mean ``shortest''; as Section~\ref{sec:ingredients} noted, if you write no
\code{#bias} at all there is no objective to minimise, so it is worth putting in even when it
looks like boilerplate.

\paragraph{The scoring function chooses the hypothesis.}
When several hypotheses fit the data, the bias decides which one is returned. The next two
examples share identical background, modes, and examples; only the \code{#bias} differs.

\exhead{Scoring, take 1: plain length}
\label{ex:bias1}
\begin{lstlisting}[style=fastlas]
#modeh(sel).
#modeb(a).
#modeb(b).

#pos(p1, {sel}, {},    { a. b. }).
#pos(p2, {},    {sel}, { c. }).

#bias("penalty(1, body(X)) :- in_body(X).").
\end{lstlisting}
\begin{lstlisting}[style=shell]
$ FastLAS --opl examples/ex06_bias_length.las
\end{lstlisting}
\begin{lstlisting}[style=result]
sel :- a.
\end{lstlisting}
Both \code{sel :- a.} and \code{sel :- b.} explain the data and both have length~1, so FastLAS
breaks the tie (here in favour of \code{a}).

\exhead{Scoring, take 2: make one literal expensive}
\label{ex:bias2}
Now we charge 10 for using \code{a} but only 1 for \code{b}:
\begin{lstlisting}[style=fastlas]
#modeh(sel).
#modeb(a).
#modeb(b).

#pos(p1, {sel}, {},    { a. b. }).
#pos(p2, {},    {sel}, { c. }).

#bias("penalty(10, body(a)) :- in_body(a).").
#bias("penalty(1,  body(b)) :- in_body(b).").
\end{lstlisting}
\begin{lstlisting}[style=shell]
$ FastLAS --opl examples/ex07_bias_custom.las
\end{lstlisting}
\begin{lstlisting}[style=result]
sel :- b.
\end{lstlisting}
The same data gives a different answer, because the scoring function is a first-class part of the specification.
This is how you encode domain-specific criteria: cost, risk, coverage, generality. Weights may be
any integers, negative ones included, which is the basis of the generality objective in the policy
experiments of \citeN{fastlas}. A second-stage \code{#final_bias("...")} lets you score
aggregate features of a whole rule after the ordinary \code{#bias} has run.

\begin{ilaspbox}
\code{#bias} means \emph{opposite things} in the two systems. In FastLAS it assigns \emph{costs}
(a scoring function). In ILASP it holds \emph{pruning constraints} (\code{:- body(...).}) that
add/remove rules from the search space but never assign cost; ILASP's objective is fixed to
hypothesis length plus example penalties. Consequently \code{#bias} lines \textbf{do not port}
between the systems.
\end{ilaspbox}

\exhead{A policy in miniature}
\label{ex:policy}
Custom scoring is why FastLAS was built: \citeN{fastlas} introduced it to learn access-control
policies from logs of granted and denied requests, where the policy that merely fits the log is
rarely the one you want. Here is the idea at its smallest. Each request is one example, its
attributes in the context; a granted request includes \code{accept}, a denied one excludes it,
and each attribute gets a \code{const} body mode:
\begin{lstlisting}[style=fastlas]
% Access-control policy learning (AAAI 2020 style): learn an `accept`
% rule from granted and denied requests. Each request's attributes live
% in that example's context.
role(manager). role(staff).   clearance(high). clearance(low).
#modeh(accept).
#modeb(subject_role(const(role))).
#modeb(subject_clearance(const(clearance))).

% granted
#pos(g1, {accept}, {},
     { subject_role(manager). subject_clearance(high). }).
% denied
#pos(d1, {}, {accept},
     { subject_role(staff). subject_clearance(low). }).

% the usual per-literal bias: "best" means "shortest"
#bias("penalty(1, head)    :- in_head(X).").
#bias("penalty(1, body(X)) :- in_body(X).").
\end{lstlisting}
\begin{lstlisting}[style=shell]
$ FastLAS --opl examples/ex18_policy.las
\end{lstlisting}
\begin{lstlisting}[style=result]
accept :- subject_role(manager).
\end{lstlisting}
Both \code{subject_role(manager)} and \code{subject_clearance(high)} are single-literal policies
consistent with the log, so under the per-literal bias they tie at cost~2 and FastLAS returns the
role-based one. Now suppose your organisation prefers \emph{clearance}-based policies; add one
line that charges for role conditions:
\begin{lstlisting}[style=fastlas]
#bias("penalty(1, prefer_clearance) :- in_body(subject_role(V)).").
\end{lstlisting}
\begin{lstlisting}[style=shell]
$ FastLAS --opl examples/ex18_policy_clearance.las
\end{lstlisting}
\begin{lstlisting}[style=result]
accept :- subject_clearance(high).
\end{lstlisting}
Same log, a different \emph{policy} chosen by the scoring criterion, not the data. The
experiments of \citeN{fastlas} push this to real access logs under length, coverage and
generality objectives; Section~\ref{sec:policy} reproduces one of them.

\paragraph{Advanced scoring: \texttt{\#final\_bias}.}
The \code{#bias} scoring of Section~\ref{sec:bias} is \emph{decomposable}: a rule's cost is a sum
of \emph{local}, per-literal contributions (\code{in_head}, \code{in_body}). That decomposability
is what lets FastLAS score candidate rules independently and stay fast. But some criteria are not
decomposable: they depend on an \emph{aggregate} property of the whole rule. The classic case:
``charge once if a rule uses negation-as-failure \emph{at all}, however many negated literals it
has.'' You cannot write that as a sum of per-literal penalties (that would count each one).

\code{#final_bias} handles these \emph{semi-decomposable} functions with a second scoring stage:
\begin{enumerate}[nosep,leftmargin=1.6em]
  \item In \code{#bias}, compute \code{intermediate/1} \emph{feature} atoms describing the whole
        rule. These may be non-local. A negated body literal appears as \code{neg(X)}, so
        \code{intermediate(naf) :- in_body(neg(X)).} makes the single atom \code{naf} a feature
        that is present \emph{iff} the rule uses any negation.
  \item \code{#final_bias} assigns the cost over those features, in a final stage after the main
        search: \code{#final_bias("penalty(1, naf) :- intermediate(naf).")}.
\end{enumerate}

\exhead{A flat cost for using negation}
\label{ex:finalbias}
The data below admits exactly one rule, \code{p :- not a, not b}, with two negated literals.
\begin{lstlisting}[style=fastlas]
% Semi-decomposable scoring: charge ONCE for using negation-as-failure,
% no matter how many negated literals a rule has. The only rule that
% fits here is
%   p :- not a, not b.
% It has two negations, yet its `naf` cost is only 1.
#modeh(p).
#modeb(not a).
#modeb(not b).

#pos(e1, {p}, {},  {}).        % nothing present -> p holds
#pos(e2, {}, {p}, { a. }).     % a present       -> p false
#pos(e3, {}, {p}, { b. }).     % b present       -> p false

% stage 1 (#bias): compute an aggregate feature of the whole rule
#bias("intermediate(naf) :- in_body(neg(X)).").
% stage 2 (#final_bias): assign the cost over those features
#final_bias("penalty(1, naf) :- intermediate(naf).").
\end{lstlisting}
\begin{lstlisting}[style=shell]
$ FastLAS --opl examples/ex16_final_bias.las
p :- not a, not b.
$ FastLAS --opl --score-only examples/ex16_final_bias.las
1
\end{lstlisting}
Its negation cost is \textbf{1}, not 2: the \code{naf} feature is charged once regardless of the
two \code{not} literals. Score the very same rule with a plain per-literal
\code{#bias("penalty(1, X) :- in_body(neg(X)).")} instead and it costs \textbf{2}, one per
negated literal. That per-literal counting is exactly the decomposable behaviour \code{#final_bias}
lets you escape.

Reach for \code{#final_bias} when your objective genuinely depends on a whole-rule property (naf
usage, the number of \emph{distinct} predicates, the presence of a particular pattern) that a
per-literal \code{#bias} cannot express. The two stages compose: \code{#bias} still does the
ordinary decomposable scoring, and \code{#final_bias} adds the aggregate part on top.


\section{Choosing the algorithm: \texttt{-{}-opl} vs.\ \texttt{-{}-nopl}}
\label{sec:oplnopl}
Almost everything so far ran under \texttt{-{}-opl}. The exceptions were the \code{num_var} runs of
Section~\ref{sec:modes}, which give the same answer either way, and the \code{#neg} task of
Section~\ref{sec:context}, which does not: that one is \code{UNSATISFIABLE} under \texttt{-{}-opl}
and solvable under \texttt{-{}-nopl}. The flag is not a cosmetic switch: it selects a
different learning algorithm, and \emph{the same \texttt{.las} file can succeed under one and
fail under the other}. There is no marker in the file itself; the choice is entirely on the
command line.

\paragraph{What \texttt{-{}-opl} assumes.}
\texttt{-{}-opl} runs the original FastLAS algorithm and does \emph{Observational Predicate
Learning}. It is valid only when, for every example:
\begin{enumerate}[nosep,leftmargin=1.6em]
  \item the predicate(s) you are learning (the \code{#modeh} heads) are \textbf{directly
        observed}, appearing in the examples' inclusions/exclusions; and
  \item $B \cup \textit{context}$ has \textbf{exactly one} answer set (a \emph{categorical}
        background: no unresolved choice rules and no even negation loops; FastLAS's parser has
        no disjunction at all).
\end{enumerate}
When these hold, \texttt{-{}-opl} is the right choice and the faster one: it skips the extra
work described next.

\paragraph{What \texttt{-{}-nopl} adds.}
\texttt{-{}-nopl} runs FastNonOPL~\cite{fastnonopl}, which inserts a
\emph{possibility-generation} phase
(abduction) before solving. This lets it learn a predicate that is \textbf{never observed} in
the examples but only influences observed predicates through the background, and it copes with
backgrounds that have \emph{multiple} answer sets. In principle anything \texttt{-{}-opl} can solve,
\texttt{-{}-nopl} can too, and the converse is false. In practice it can also be substantially
slower. Reach for \texttt{-{}-nopl} when you need it, not by default.

\exhead{A non-observational target: \texttt{-{}-opl} fails, \texttt{-{}-nopl} succeeds}
\label{ex:nopl}
We observe \code{fault}; we want to learn \code{permitted}, which appears only inside the
background rule \code{fault :- did(A), not permitted(A), act(A)}, never in an example.
\begin{lstlisting}[style=fastlas]
fault :- did(A), not permitted(A), act(A).

#modeh(permitted(var(act))).
#modeb(authorised(var(act))).
#maxv(1).
#bias("penalty(1, X) :- in_head(X).").
#bias("penalty(1, X) :- in_body(X).").

#pos(e1, {},      {fault}, { act(a1). did(a1). authorised(a1). }).
#pos(e2, {fault}, {},      { act(a2). did(a2). }).
\end{lstlisting}
\begin{lstlisting}[style=shell]
$ FastLAS --opl  examples/ex08_nopl_permitted.las
\end{lstlisting}
\begin{lstlisting}[style=result]
UNSATISFIABLE
\end{lstlisting}
\begin{lstlisting}[style=shell]
$ FastLAS --nopl examples/ex08_nopl_permitted.las
\end{lstlisting}
\begin{lstlisting}[style=result]
permitted(V0) :- authorised(V0), act(V0).
\end{lstlisting}
Under \texttt{-{}-opl} the only learnable head is \code{permitted}, but the \emph{observed} atom
is \code{fault}; FastLAS has no way to connect them, so no hypothesis covers the examples. Under
\texttt{-{}-nopl}, possibility generation reasons backwards through the background rule
\code{fault :- ..., not permitted(A), ...} to infer what \code{permitted} must look like, and
recovers the target rule.

\exhead{Non-observational learning on a real task}
\label{ex:agent}
This task learns \code{valid_move}, which appears only in the background rule
\code{violation :- agent_at(C,T), not valid_move(C,T-1), time(T-1)}, from examples that mention
only \code{violation}:
\begin{lstlisting}[style=fastlas]
#pos(eg3, {}, {violation}, {
  agent_at(cell(10,1),1).
  agent_at(cell(10,2),2).
  agent_at(cell(10,1),3).
  agent_at(cell(9,1),4).
  agent_at(cell(8,1),5).
}).

#pos(eg0, {violation}, {}, {
  agent_at(cell(10,1),1).
  agent_at(cell(10,2),2).
  agent_at(cell(8,1),3).
}).


cell(cell(1..10, 1..10)).
adjacent(cell(X,Y),cell(X+1,Y)) :- cell(cell(X,Y)), cell(cell(X+1,Y)).
adjacent(cell(X+1,Y),cell(X,Y)) :- cell(cell(X,Y)), cell(cell(X+1,Y)).
adjacent(cell(X,Y+1),cell(X,Y)) :- cell(cell(X,Y)), cell(cell(X,Y+1)).
adjacent(cell(X,Y),cell(X,Y+1)) :- cell(cell(X,Y)), cell(cell(X,Y+1)).

violation :- agent_at(C, T), not valid_move(C, T-1), time(T-1).

leq(T, T2) :- time(T), time(T2), T <= T2.
time(1..5).

#modeh(valid_move(var(cell), var(time))).
#modeb(agent_at(var(cell), var(time))).
#modeb(adjacent(var(cell), var(cell))).
#bias("penalty(1, head(X)) :- in_head(X).").
#bias("penalty(1, body(X)) :- in_body(X).").
\end{lstlisting}
\begin{lstlisting}[style=shell]
$ FastLAS --opl examples/ex09_agent_abduction.las
\end{lstlisting}
\begin{lstlisting}[style=result]
UNSATISFIABLE
\end{lstlisting}
\begin{lstlisting}[style=shell]
$ FastLAS --nopl examples/ex09_agent_abduction.las
\end{lstlisting}
\begin{lstlisting}[style=result]
valid_move(V0,V1) :- agent_at(V2,V1), adjacent(V2,V0), cell(V0),
        time(V1), cell(V2).
\end{lstlisting}

\begin{oplbox}
\textbf{Decision rule.} Use \texttt{-{}-opl} when your target predicates are directly observed
and the background is deterministic (one answer set per context); it is faster. Switch to
\texttt{-{}-nopl} when either (a) the predicate you are learning never appears in the examples
(non-observational), or (b) the background/context can have more than one answer set, which in
FastLAS comes from a choice rule or from an even loop through negation such as
\code{a :- not b.} with \code{b :- not a.} A symptom of needing \texttt{-{}-nopl} is an
\texttt{-{}-opl} run that
returns \texttt{UNSATISFIABLE} even though you believe a rule exists.
\end{oplbox}

\begin{center}
\begin{tabular}{@{}lll@{}}
\toprule
 & \texttt{-{}-opl} & \texttt{-{}-nopl} \\
\midrule
Algorithm                       & original FastLAS  & FastNonOPL \\
Target must be observed?         & yes               & no \\
Background may have $>1$ answer set? & no            & yes \\
Extra abduction phase            & no                & yes \\
Speed                            & faster            & slower \\
Generality                       & subset            & superset (in principle) \\
\bottomrule
\end{tabular}
\end{center}

\section{Writing effective programs}
\label{sec:effective}
Beyond getting a task to run, four topics recur in practice: keeping the search fast, working within FastLAS's expressiveness limits, querying a learned theory, and learning from a stream.

\subsection{Writing an efficient mode bias}
\label{sec:perf}
The mode declarations decide both \emph{what} FastLAS can learn and \emph{how long it takes}.
FastLAS builds candidate rules by grounding the modes over variable assignments, and that work
grows combinatorially with the number of variables you allow (\code{#maxv}) and the number and
kind of body modes. The catch is that this cost is paid \emph{even when the final hypothesis is
tiny}: a bias that is looser than the target needs, but still contains it, gives the same answer more
slowly. Two rules of thumb
dominate.

\paragraph{Keep \texttt{\#maxv} as small as the target needs.}
The bound on distinct variables per rule is the strongest lever you have on the size of the
search, as the following experiment shows.

\exhead{A performance experiment (tightening the variable bound)}
\label{ex:perf}
This task learns \code{rel(V0,V1) :- edge(V1,V0)}, a target that needs only two
variables.
\begin{lstlisting}[style=fastlas]
node(a). node(b). node(c). node(d). node(e). node(f). node(g). node(h).
edge(a,b). edge(b,a). edge(b,c). edge(c,b). edge(c,d). edge(d,c).
edge(e,f). edge(f,e). edge(f,g). edge(g,f). edge(g,h). edge(h,g).
near(a,c). near(c,a). near(e,g). near(g,e).
far(a,h). far(h,a). far(a,g). far(d,e).

#modeh(rel(var(node), var(node))).
#modeb(edge(var(node), var(node))).
#modeb(near(var(node), var(node))).
#modeb(far(var(node), var(node))).
#modeb(var(node) != var(node)).

#pos(p1, {rel(a,b)}, {},        { edge(a,b). }).
#pos(p2, {rel(c,d)}, {},        { edge(c,d). }).
#pos(p3, {rel(f,g)}, {},        { edge(f,g). }).
#pos(p4, {},         {rel(a,c)}, { near(a,c). }).
#pos(p5, {},         {rel(a,h)}, { far(a,h). }).

#bias("penalty(1, head) :- in_head(X).").
#bias("penalty(1, body(X)) :- in_body(X).").
#maxv(2).
\end{lstlisting}
\begin{lstlisting}[style=shell]
$ FastLAS --opl --space-size examples/ex12_modebias.las
% SPACE SIZE: 1
rel(V0,V1) :- edge(V1,V0), node(V0), node(V1).
\end{lstlisting}
It learns the intended rule already at \code{#maxv(2)}; \code{#maxv} is a directive rather than a
flag, so to reproduce the rows below, copy the file and edit its last line. Watch what happens to
the running time
as we raise \code{#maxv}. The times below were measured on a MacBook Pro (13-inch, 2020,
model \texttt{MacBookPro17,1}) with an Apple~M1 system-on-chip (8 cores: 4 performance and 4
efficiency) and 16\,GB of unified memory, using FastLAS 2.2.0 and Clingo 5.8.0; absolute times
will differ on other machines, and what matters is how sharply they grow:
\begin{center}
\begin{tabular}{@{}cccl@{}}
\toprule
\code{#maxv} & learned rule & final space size & wall-clock time \\
\midrule
2 & \code{rel(V0,V1):-edge(V1,V0)} & 1 & 0.06\,s \\
3 & (same) & 2 & 0.09\,s \\
4 & (same) & 2 & 1.7\,s \\
5 & --- & --- & killed after 2\,min \\
\bottomrule
\end{tabular}
\end{center}
Neither the learned hypothesis nor the final hypothesis-space size FastLAS reports
(\code{SPACE SIZE}, the number of candidate rules FastLAS keeps after pruning) grows past two rules, yet raising
\code{#maxv} from 2 to 5 turns a $0.06$-second run into one that does not finish in two minutes.
Every extra permitted variable multiplies the assignments FastLAS must ground. \textbf{Set
\texttt{\#maxv} to the smallest value in which your intended rule can be written.}

\paragraph{Declare only the body modes you need.}
Extra body modes enlarge the same grounding. \emph{Comparison} modes such as
\code{var(t) != var(t)} or \code{var(t) < var(t)} are the worst offenders, because they range
over \emph{every pair} of the allowed variables. On the same task, at a fixed \code{#maxv(4)}, growing from one body mode to three takes
$0.09$\,s to $0.44$\,s, and adding the single \code{!=} comparison as a fourth mode takes it to
$1.74$\,s: a further factor of four for one declaration, with no change to the learned rule.

\begin{gotcha}
\textbf{Symptoms of an over-loose bias:} a run that takes far longer than the size of the answer
would suggest, or that appears to hang. Tighten it by (1)~lowering \code{#maxv}; (2)~removing modes the target does not use,
especially comparison and high-arity modes; and (3)~using \code{const(t)} instead of
\code{var(t)} wherever a fixed value suffices. \code{--space-size} reports the final space size
and \code{--debug} shows which phase is slow.
\end{gotcha}

\subsection{Working within FastLAS's limits}
\label{sec:limits}
FastLAS learns \emph{non-recursive} normal rules; constraint-style checks are learned through a
fixed head such as \code{violated}, as below. Under \code{--opl} the
background must also be \emph{categorical}, one answer set per example; \code{--nopl} lifts that
particular restriction (Section~\ref{sec:oplnopl}), but neither algorithm learns recursion. Three practical
consequences follow, each with an idiom that works around it.

\paragraph{Recursion belongs in the background.}
A \emph{learned} rule may not be recursive, but the background may. Precompute any recursive
feature (reachability, transitive closure, \dots) in the background and expose it to the learner
as an ordinary body mode.

\exhead{A recursively-defined feature in the background}
\label{ex:recursion}
\begin{lstlisting}[style=fastlas]
% Recursion is allowed in the BACKGROUND: reach/2 is a transitive
% closure, while the LEARNED rule stays non-recursive and just uses the
% derived feature connected/0.
reach(X,Y) :- edge(X,Y).
reach(X,Z) :- edge(X,Y), reach(Y,Z).
connected  :- reach(a,d).

#modeh(target).
#modeb(connected).

% path a..d -> connected
#pos(p1, {target}, {},       { edge(a,b). edge(b,c). edge(c,d). }).
% gap -> not connected
#pos(n1, {},       {target}, { edge(a,b). edge(c,d). }).

#bias("penalty(1, head)    :- in_head(X).").
#bias("penalty(1, body(X)) :- in_body(X).").
\end{lstlisting}
\begin{lstlisting}[style=shell]
$ FastLAS --opl examples/ex13_recursion_bg.las
\end{lstlisting}
\begin{lstlisting}[style=result]
target :- connected.
\end{lstlisting}
\code{reach/2} is defined recursively in the background; the learned rule
\code{target :- connected.} is non-recursive and merely \emph{uses} the reachability feature.

\paragraph{Reframe a coupled target as a verifier.}
The running example here is \emph{abstract argumentation}~\cite{dung}, and one paragraph of it
is all we need. An argumentation framework is a directed graph: the nodes are arguments, and an
edge \code{att(X,Y)} says that \code{X} attacks \code{Y}. A \emph{labelling} marks each
argument \code{in} (accepted, drawn with a double border) or \code{out} (rejected, dashed), and
the basic sanity condition, \emph{conflict-freeness}, says that no \code{in} argument may attack
another \code{in} argument.
\begin{center}
\begin{tikzpicture}[>={Stealth[round]},
    every node/.style={circle,draw,inner sep=1.6pt,font=\small}]
  \node[double]  (a) at (0,0)   {$a$};
  \node[dashed]  (b) at (1.6,0) {$b$};
  \node[double]  (c) at (3.2,0) {$c$};
  \draw[->] (a) -- (b);
  \draw[->] (b) -- (c);
\end{tikzpicture}
\end{center}
In the framework drawn above, the labelling shown (\code{a} and \code{c} in, \code{b} out) is
conflict-free; labelling \code{a} and \code{b} both \code{in} is not, because \code{a} attacks
\code{b}.

Here is the trap. Suppose you try to learn such a labelling
\code{in}/\code{out} directly, and the
feature that would discriminate, ``\code{X} is attacked by an \code{in} argument'', is
itself defined from \code{in}. That asks FastLAS for a \emph{recursive} rule (the target appears
in the definition of its own body feature), which it will not produce: the direct task either
returns \code{UNSATISFIABLE} or collapses to a trivial over-general rule.

The fix is the \emph{verifier reframing}: give the labelling as \textbf{context} (so the feature
becomes deterministic \emph{input}, not something derived from the target), and learn a
\emph{constraint}, a single \code{violated} head (or \code{#modeh(false)}), that recognises
the bad cases. Valid examples \emph{exclude} \code{violated}; invalid ones \emph{include} it.

\exhead{Learning a verifier over a given labelling}
\label{ex:verifier}
\begin{lstlisting}[style=fastlas]
% Verifier reframing: the labelling in/1 and out/1 is GIVEN as context,
% so attacked_by_in/1 is deterministic input rather than something
% learned through the target predicate.
% We learn a constraint-style verifier:
%   violated :- <features>.
arg(a).  arg(b).
attacked_by_in(X) :- att(Y,X), in(Y).

#modeh(violated).
#modeb(in(var(arg))).
#modeb(out(var(arg))).
#modeb(attacked_by_in(var(arg))).
#modeb(not attacked_by_in(var(arg))).
#maxv(1).

% valid   -> not violated
#pos(cf1,  {},          {violated}, { att(a,b). in(a). out(b). }).
% valid   -> not violated
#pos(cf2,  {},          {violated}, { att(a,b). out(a). in(b). }).
% invalid -> violated
#pos(bad1, {violated},  {},         { att(a,b). in(a). in(b). }).

#bias("penalty(1, body(X)) :- in_body(X).").
\end{lstlisting}
\begin{lstlisting}[style=shell]
$ FastLAS --opl examples/ex14_verifier.las
\end{lstlisting}
\begin{lstlisting}[style=result]
violated :- in(V0), attacked_by_in(V0), arg(V0).
\end{lstlisting}
FastLAS learns the textbook conflict-freeness check: a labelling is bad when an \code{in} argument
is attacked by an \code{in} argument. Because the labelling is context (deterministic, no choice
rules), this runs under the faster \code{--opl}.

\paragraph{Rebuild the full semantics at deploy time.}
A verifier only \emph{checks} labellings; to \emph{enumerate} them, pair the learned rule with a
fixed \emph{generate} step in a plain Clingo program. Everything here is hand-written except the
one learned line:
\begin{lstlisting}[style=fastlas]
% Deploying a learned verifier (clingo, not FastLAS): a fixed generate
% step plus the rule FastLAS learned in the verifier example, turned
% into a constraint. clingo then enumerates the conflict-free
% labellings of the framework.
arg(a). arg(b). att(a,b).
0 { in(X) } 1 :- arg(X).
out(X) :- arg(X), not in(X).
attacked_by_in(X) :- att(Y,X), in(Y).
violated :- in(V0), attacked_by_in(V0), arg(V0).   % learned
:- violated.
#show in/1.
\end{lstlisting}
\begin{lstlisting}[style=shell]
$ clingo examples/asp/asp10_deploy.lp --models=0
Answer: 1

Answer: 2
in(b)
Answer: 3
in(a)
SATISFIABLE
\end{lstlisting}
Exactly the three conflict-free labellings of the one-attack framework, no more. Any \emph{global}
objective, preferring larger labellings with a fixed \code{#maximize}, say, also lives in this
deploy program rather than in the learning task: FastLAS learns the local \emph{check}, and the
fixed part provides the rest.

\begin{oplbox}
This trap is not the non-observational case of Section~\ref{sec:oplnopl}: there the target never
appears in the examples, whereas here it is exactly what the examples record. The obstacle is
recursion, the target appearing in the definition of its own body feature, and neither
\code{--opl} nor \code{--nopl} learns recursion, so the direct task stays
\code{UNSATISFIABLE} under both. The verifier reframing is the remedy, and it keeps you in the
faster \code{--opl}.
\end{oplbox}

\subsection{Prediction queries: \texttt{\#predict}}
\label{sec:predict}
Sometimes you do not want the hypothesis itself but an answer to a question: \emph{given what
FastLAS would learn, does the theory predict \textnormal{X} in situation \textnormal{Y}?} The
\code{#predict} directive asks exactly that. Syntactically it mirrors an example, a partial
interpretation plus a context,
\begin{center}
\begin{lstlisting}[style=fastlas]
#predict( id , { inclusions } , { exclusions } , { context } ).
\end{lstlisting}
\end{center}
but instead of constraining learning, it \emph{queries} the learned theory. At most one
\code{#predict} may appear in a task.

\paragraph{What FastLAS prints.}
Adding a \code{#predict} changes the output. Instead of one hypothesis, FastLAS prints
\textbf{two}, each followed by a small JSON cost summary:
\begin{itemize}[nosep,leftmargin=1.4em]
  \item the cheapest hypothesis (consistent with all the examples) that \emph{satisfies} the
        query, and
  \item the cheapest one that does \emph{not} satisfy it.
\end{itemize}
Reading off the verdict is easy:
\begin{itemize}[nosep,leftmargin=1.4em]
  \item if the \emph{not-satisfying} side is \texttt{UNSATISFIABLE}, the query is
        \textbf{necessarily entailed}: no admissible theory can avoid it (a confident yes);
  \item if the \emph{satisfying} side is \texttt{UNSATISFIABLE}, the query is
        \textbf{impossible} (a confident no);
  \item if both sides return a hypothesis, compare their \code{Length}: the optimal theory
        behaves like the cheaper side (equal cost means the data leaves the query undetermined).
\end{itemize}

\exhead{A prediction that is necessarily entailed}
\label{ex:predict-yes}
We reuse the cycling task and ask: on a clear day (empty context), does the theory predict
\code{cycle}?
\begin{lstlisting}[style=fastlas]
#modeh(cycle).
#modeb(rain).
#modeb(not rain).

#pos(d1, {cycle}, {}, {}).
#pos(d2, {}, {cycle}, {rain.}).

#bias("penalty(1, head) :- in_head(X).").
#bias("penalty(1, body(X)) :- in_body(X).").

#predict(clear_day, {cycle}, {}, { }).
\end{lstlisting}
\begin{lstlisting}[style=shell]
$ FastLAS --opl examples/ex10_predict_entailed.las
\end{lstlisting}
\begin{lstlisting}[style=result]
% Optimal hypothesis satisfying the prediction:

cycle :- not rain.

{
  "Length": 2,
  "Noisy Example Penalty": 0,
  "Uncovered Examples": [ ],
  "Final Semi-decomposable Representation": [ ]
}

% Optimal hypothesis not satisfying the prediction:

UNSATISFIABLE

{
  "Length": 0, ...
}
\end{lstlisting}
The not-satisfying side is \texttt{UNSATISFIABLE}: \emph{no} hypothesis consistent with the
training data can make \code{cycle} false on a clear day, so the prediction is a confident yes.

\exhead{A prediction that is impossible}
\label{ex:predict-no}
The mirror-image question: on a rainy day, does the theory predict \code{cycle}? Only the
context changes (\code{{rain.}}).
\begin{lstlisting}[style=fastlas]
#modeh(cycle).
#modeb(rain).
#modeb(not rain).

#pos(d1, {cycle}, {}, {}).
#pos(d2, {}, {cycle}, {rain.}).

#bias("penalty(1, head) :- in_head(X).").
#bias("penalty(1, body(X)) :- in_body(X).").

#predict(rainy_day, {cycle}, {}, { rain. }).
\end{lstlisting}
\begin{lstlisting}[style=shell]
$ FastLAS --opl examples/ex11_predict_impossible.las
\end{lstlisting}
\begin{lstlisting}[style=result]
% Optimal hypothesis satisfying the prediction:

UNSATISFIABLE

{ "Length": 0, ... }

% Optimal hypothesis not satisfying the prediction:

cycle :- not rain.

{ "Length": 2, ... }
\end{lstlisting}
Now the \emph{satisfying} side is \texttt{UNSATISFIABLE}: no admissible theory makes
\code{cycle} hold when it rains, a confident no. (When both sides come back with a
hypothesis, you instead compare the two \code{Length} values; the FastLAS source distribution
ships such a task as \code{FastLAS2/testing/prediction_task.las}.)

\subsection{Incremental learning from a stream (the cache)}
\label{sec:incremental}
When examples arrive over time, a \emph{stream} of \emph{windows} rather than one fixed
training set, you do not want to redo all the work each time a new window shows up. FastLAS can
cache the hypothesis space and the per-example analysis computed so far and reload it when the next
window arrives (the IncrementalLAS algorithm of \citeNP{incrementallas}). The guarantee: after each
window the result is the
\emph{same} optimal hypothesis you would get by learning from every window at once. Two flags drive
it:
\begin{center}
\begin{tabular}{@{}l p{0.6\textwidth}@{}}
\toprule
\code{--write-cache FILE} & after solving, write the computed state (space + per-example data) to \code{FILE} \\
\code{--read-cache FILE}  & resume from a previously-written cache instead of recomputing it \\
\bottomrule
\end{tabular}
\end{center}

\exhead{A two-window stream}
\label{ex:incremental}
Each window file carries the fixed parts (mode bias, \code{#bias}, background) plus only that
window's \emph{new} examples. Window~1 has two examples; window~2 adds a single new one:
\begin{lstlisting}[style=fastlas]
% fixed parts (repeated every window): mode bias + scoring
#modeh(p).
#modeb(a).
#modeb(b).
#bias("penalty(1, body(X)) :- in_body(X).").

% window 1 examples
#pos(e1, {p}, {},  { a. b. }).
#pos(e2, {}, {p}, { }).
\end{lstlisting}
\begin{lstlisting}[style=fastlas]
% fixed parts (repeated every window): mode bias + scoring
#modeh(p).
#modeb(a).
#modeb(b).
#bias("penalty(1, body(X)) :- in_body(X).").

% window 2 brings ONE new example (the cache carries e1, e2)
#pos(e3, {}, {p}, { a. }).
\end{lstlisting}
Learn window~1 and save its cache; then learn window~2 \emph{reading} that cache and saving a new one:
\begin{lstlisting}[style=shell]
$ FastLAS --opl --write-cache cache1 examples/ex15_window1.las
\end{lstlisting}
\begin{lstlisting}[style=result]
p :- a.
\end{lstlisting}
\begin{lstlisting}[style=shell]
$ FastLAS --opl --read-cache cache1 --write-cache cache2 \
      examples/ex15_window2.las
\end{lstlisting}
\begin{lstlisting}[style=result]
p :- b.
\end{lstlisting}
The hypothesis \emph{refined} from \code{p :- a.} to \code{p :- b.} once the new example ruled out
the first rule. Window~2 was passed \emph{only} the new example \code{e3}; \code{e1} and
\code{e2} came from the cache. Chain further windows the same way
(\code{--read-cache cache2 --write-cache cache3 ...}).

\textbf{The guarantee.} Learning the same three examples in one batch gives the identical result:
\begin{lstlisting}[style=shell]
$ FastLAS --opl examples/ex15_batch.las          # e1, e2, e3 together
\end{lstlisting}
\begin{lstlisting}[style=result]
p :- b.
\end{lstlisting}
Incremental equals batch: the cache saves work, not accuracy.

\begin{gotcha}
Each window file must still contain the \code{#modeh}/\code{#modeb}/\code{#bias} declarations and
any global background. The cache carries the computed \emph{per-example} data and the
hypothesis-space schemas, but the mode bias must be present for FastLAS to fold the new window's
examples in; drop it and the new examples are silently ignored (you get the previous window's
answer). Treat the cache as opaque. Do not hand-edit it.
\end{gotcha}

The payoff is what makes streaming tractable: each window reuses the earlier windows' analysis
instead of recomputing it, while the guarantee above keeps the running hypothesis optimal
over everything seen so far.


\section{Classic problems, from Clingo to FastLAS}
\label{sec:classics}
The problems in this section are the standard exercises of an ASP course: cliques, colouring,
vertex cover, and a small preference puzzle. Each one can be read twice. Written for Clingo,
the program \emph{solves} an instance, and you supply the rules. Written for FastLAS, the rules are
what you are missing, and the instances are what you supply. Several of the encodings below go back to the
textbook treatments of \citeN{lifschitz} and \citeN{clingo}.

The Clingo programs below all use the same six-vertex graph (the learning tasks later in the
section use smaller ones, so that the examples stay readable):
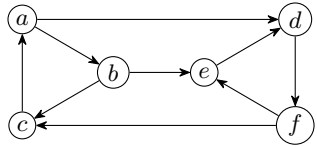
\begin{center}
\begin{tikzpicture}[>={Stealth[round]}, every node/.style={circle,draw,inner sep=1.6pt,
                    font=\small}]
  \node (a) at (0,1.4)   {$a$};
  \node (d) at (3.6,1.4) {$d$};
  \node (b) at (1.2,0.7) {$b$};
  \node (e) at (2.4,0.7) {$e$};
  \node (c) at (0,0)     {$c$};
  \node (f) at (3.6,0)   {$f$};
  \draw[->] (a) -- (b);  \draw[->] (b) -- (c);  \draw[->] (c) -- (a);
  \draw[->] (d) -- (f);  \draw[->] (f) -- (e);  \draw[->] (e) -- (d);
  \draw[->] (a) -- (d);  \draw[->] (f) -- (c);  \draw[->] (b) -- (e);
\end{tikzpicture}
\end{center}
\begin{lstlisting}[style=fastlas]
vertex(a;b;c;d;e;f).
edge(a,b; b,c; c,a; d,f; f,e; e,d; a,d; f,c; b,e).
\end{lstlisting}

\subsection{Guess and test in Clingo}
\label{sec:guessandtest}
The Clingo idiom for these problems is \emph{guess and test}: a choice rule guesses a candidate,
and constraints reject the candidates that are not solutions. A clique is one line of each.
\begin{lstlisting}[style=fastlas]
{ in(V) : vertex(V) } = 3.
:- in(V1), in(V2), V1!=V2, not edge(V1,V2), not edge(V2,V1).
\end{lstlisting}
\begin{lstlisting}[style=shell]
$ clingo examples/asp/asp02_clique.lp
in(e) in(f) in(d)
\end{lstlisting}
Vertex cover is the same shape, and introduces the auxiliary predicate that the learning task
below will have to discover:
\begin{lstlisting}[style=fastlas]
{in(X) : vertex(X)}.
covered(X,Y) :- edge(X,Y), in(X).
covered(X,Y) :- edge(X,Y), in(Y).
:- edge(X,Y), not covered(X,Y).
\end{lstlisting}
\begin{lstlisting}[style=shell]
$ clingo examples/asp/asp05_vertexcover.lp
in(b) in(c) in(d) in(e)
\end{lstlisting}
None of this is FastLAS input, starting with the very first line: the pooling semicolon of
\code{vertex(a;b;c;d;e;f).} is a syntax error there, and so are a choice rule whose head begins
with a left brace (FastLAS wants both bounds), conditional literals with \code{:}, and
\code{#show}. All are outside the ASP dialect it accepts (Section~\ref{sec:bg}); these
programs are for Clingo, and they are in \texttt{examples/asp/}.

Two more classics ship in \code{examples/asp/} for reading rather than walking through:
\code{asp04_queens.lp}, and \code{asp03_hamiltonian.lp}, which carries a caveat worth the detour.
Its reachability rule follows \code{edge/2} rather than the guessed \code{in/2}, so the test is
vacuous: the file admits 440 models, where the corrected version (one literal changed to
\code{in(V1,V)}) admits 6, the same single cycle counted once per guessed start vertex.
Hand-written encodings are delicate; learning the rule from examples, as the tasks below do, is
one way of not hand-writing them.

\subsection{The same problems as learning tasks}
\label{sec:classiclearning}
Turn each program around. Instead of writing \code{covered/2} and asking Clingo for a cover, give
FastLAS some graphs in which you already know what is covered, and let it write the rule.

\exhead{Learning the vertex-cover rule}
\label{ex:vertexcover}
The graph and the guessed set \code{in} become example contexts, and the head is the predicate
we want defined.
\begin{lstlisting}[style=fastlas]
% Learn one half of the vertex-cover definition:
%   covered(X,Y) :- edge(X,Y), in(X).
% The graph and the guessed set "in" are supplied as example contexts.
vertex(a). vertex(b).
#modeh(covered(var(vertex), var(vertex))).
#modeb(edge(var(vertex), var(vertex))).
#modeb(in(var(vertex))).
#maxv(2).
#pos(p1, {covered(a,b)}, {},             { edge(a,b). in(a). }).
#pos(p2, {},             {covered(a,b)}, { edge(a,b). in(b). }).
#pos(p3, {},             {covered(a,b)}, { edge(a,b).        }).
% in(a) but no edge
#pos(p4, {},             {covered(a,b)}, { in(a).            }).
#bias("penalty(1, head)    :- in_head(X).").
#bias("penalty(1, body(X)) :- in_body(X).").
\end{lstlisting}
\begin{lstlisting}[style=shell]
$ FastLAS --opl examples/ex19_vertexcover.las
covered(V0,V1) :- edge(V0,V1), in(V0), vertex(V0), vertex(V1).
\end{lstlisting}
That is the first of the two rules above. Example \code{p4} is the one that earns its keep. Drop
it and the answer degrades to \code{covered(V0,V1) :- in(V0), vertex(V0), vertex(V1).}, which
explains every remaining example and is one literal shorter, because nothing in the task ever
shows a vertex that is \code{in} without an edge leaving it. It is the concrete form of the rule
in Section~\ref{sec:context}: an example is only informative if some candidate rule gets it
wrong.

\exhead{Learning the colouring conflict}
\label{ex:colouring}
A constraint is learned as a rule with a fixed head, the verifier reframing of
Section~\ref{sec:limits}. Proper colourings exclude \code{violated}, improper ones include it.
\begin{lstlisting}[style=fastlas]
% Learn the graph-colouring conflict as a verifier:
%   violated :- edge(V0,V1), colour(V0,V2), colour(V1,V2).
vertex(a). vertex(b).  shade(red). shade(blue).
#modeh(violated).
#modeb(edge(var(vertex), var(vertex))).
#modeb(colour(var(vertex), var(shade))).
#maxv(3).
#pos(ok1, {}, {violated},
     { edge(a,b). colour(a,red). colour(b,blue). }).
#pos(ok2, {}, {violated},
     { edge(a,b). colour(a,blue). colour(b,red). }).
#pos(bad1, {violated}, {},
     { edge(a,b). colour(a,red). colour(b,red). }).
#pos(bad2, {violated}, {},
     { edge(a,b). colour(a,blue). colour(b,blue). }).
#bias("penalty(1, body(X)) :- in_body(X).").
\end{lstlisting}
\begin{lstlisting}[style=shell]
$ FastLAS --opl examples/ex20_colouring.las
violated :- edge(V1,V2), colour(V2,V0), colour(V1,V0), shade(V0),
        vertex(V1), vertex(V2).
\end{lstlisting}
Two vertices joined by an edge carrying the same shade \code{V0}: the Clingo constraint
\code{:- edge(X,Y), colour(X,C1), colour(Y,C2), C1 = C2.} recovered from examples alone.

\exhead{Learning the clique conflict}
\label{ex:clique}
The same reframing recovers the clique test of Section~\ref{sec:guessandtest}. The graph is the
three-vertex one below, with a single edge, and each context pairs it with one chosen set:
$\{a,b\}$ and the singleton $\{a\}$ are cliques, $\{a,c\}$ and $\{b,c\}$ are not.
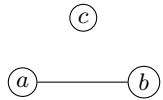
\begin{center}
\begin{tikzpicture}[every node/.style={circle,draw,inner sep=1.6pt,font=\small}]
  \node (a) at (0,0)   {$a$};
  \node (b) at (1.6,0) {$b$};
  \node (c) at (0.8,0.85) {$c$};
  \draw (a) -- (b);
\end{tikzpicture}
\end{center}
The background makes the adjacency symmetric, so one negated literal is enough to say ``not
neighbours'':
\begin{lstlisting}[style=fastlas]
% Learning the clique conflict: a chosen set is bad as soon as it holds
% two distinct vertices that are not adjacent. The background makes the
% adjacency symmetric so one literal suffices.
vertex(a). vertex(b). vertex(c).
adj(X,Y) :- edge(X,Y).
adj(X,Y) :- edge(Y,X).

#modeh(violated).
#modeb(in(var(vertex))).
#modeb(not adj(var(vertex), var(vertex))).
#modeb(var(vertex) != var(vertex)).
#maxv(2).

#pos(ok1,  {},         {violated}, { edge(a,b). in(a). in(b). }).
#pos(ok2,  {},         {violated}, { edge(a,b). in(a). }).
#pos(bad1, {violated}, {},         { edge(a,b). in(a). in(c). }).
#pos(bad2, {violated}, {},         { edge(a,b). in(b). in(c). }).

#bias("penalty(1, body(X)) :- in_body(X).").
\end{lstlisting}
\begin{lstlisting}[style=shell]
$ FastLAS --opl examples/ex33_clique.las
violated :- in(V0), in(V1), not adj(V1,V0), V1 != V0, vertex(V0),
        vertex(V1).
\end{lstlisting}
Two chosen, distinct, non-adjacent vertices: the clique constraint of
Section~\ref{sec:guessandtest}, recovered with the comparison mode \code{var(t) != var(t)} of
Section~\ref{sec:modes} doing the ``distinct'' part.

\begin{gotcha}
The learned rule is stable, but \textbf{the order of its body literals is not}. Across repeated
runs of the same task, the colouring rule above comes back with \code{edge/2} first or with a
\code{colour/2} literal first, and the pet-owner rule of Example~\ref{ex:petowner} alternates
between \code{eats(V0,V1)} and \code{eats(V1,V0)} with the two \code{own/1} literals swapped to
match. These are the same rule up to the order of a conjunction and the naming of variables.
Compare hypotheses as sets of literals, not as strings, and do not build a regression test that
diffs the output character by character.
\end{gotcha}

\subsection{Preferences: where FastLAS stops and ILASP starts}
\label{sec:preferences}
The pet-owner puzzle has three animals, and cats eat fish while dogs eat cats.
A farmer wants as many animals as possible, and would rather they did not eat one another. In
Clingo those are two weak constraints at different priorities. The no-eating one carries the
higher priority \code{@2}, so Clingo satisfies it first and sacrifices animals to it:
\begin{lstlisting}[style=fastlas]
0 { own(A) } 1 :- animal(A).
:~ own(A).[-1@1, A]
:~ own(A), own(B), eats(A, B).[1@2, A, B]
\end{lstlisting}
\begin{lstlisting}[style=shell]
$ clingo examples/asp/asp08_petowner.lp
own(fish) own(dog)
Optimization: 0 -2
\end{lstlisting}
Two animals, no conflicts. FastLAS cannot learn that pair of weak constraints, and cannot learn
the choice rule either. What it can learn is the conflict itself, as a hard rule.

\exhead{Learning the pet-owner conflict}
\label{ex:petowner}
\begin{lstlisting}[style=fastlas]
% The pet-owner problem. Clingo scores the conflict
% with a weak constraint; FastLAS learns the conflict itself as a hard
% rule:
%   violated :- own(V0), own(V1), eats(V0,V1).
animal(dog). animal(cat). animal(fish).
#modeh(violated).
#modeb(own(var(animal))).
#modeb(eats(var(animal), var(animal))).
#maxv(2).
#pos(ok1, {}, {violated},
     { own(dog). own(fish). eats(cat,fish). eats(dog,cat). }).
#pos(ok2, {}, {violated}, { own(cat). eats(cat,fish). eats(dog,cat). }).
#pos(bad1, {violated}, {},
     { own(cat). own(fish). eats(cat,fish). eats(dog,cat). }).
#pos(bad2, {violated}, {},
     { own(dog). own(cat). eats(cat,fish). eats(dog,cat). }).
#bias("penalty(1, body(X)) :- in_body(X).").
\end{lstlisting}
\begin{lstlisting}[style=shell]
$ FastLAS --opl examples/ex21_petowner.las
violated :- own(V0), own(V1), eats(V0,V1), animal(V0), animal(V1).
\end{lstlisting}

\begin{ilaspbox}
Preferences are the sharpest line between the two systems. ILASP learns weak constraints
directly: you label examples and order them with \code{#brave_ordering(id, eg1, eg2).} and
\code{#cautious_ordering(...)}, and it returns the \code{:~} rules that reproduce the ordering.
This is what makes it usable for preference learning on data such as the SUSHI rankings of
\citeN{sushi}, where each observation is one item preferred to another. FastLAS has no ordering
examples and no \code{#modeo}, so an ordering task has to be reformulated as a classification
task before FastLAS can take it, as in Example~\ref{ex:sushi}.
\end{ilaspbox}

\exhead{A SUSHI taste, as a classification task}
\label{ex:sushi}
The SUSHI data of \citeN{sushi} describes each item by style, major and minor group, oiliness,
price and frequency. Rather than an ordering over items, ask which items a user likes, and let
the numeric features find their own thresholds with \code{num_var}
(Section~\ref{sec:modes}).
\begin{lstlisting}[style=fastlas]
% A SUSHI-style preference task using the feature encoding (style, major group, oiliness, price; cf. Kamishima's SUSHI
% preference data).
% The taste to be recovered: seafood that is oily enough.
%   likes(S) :- oiliness(S,O), O >= 3, seafood(S).
sushi(s1). sushi(s2). sushi(s3). sushi(s4). sushi(s5). sushi(s6).
oiliness_val(0..4).

#modeh(likes(var(sushi))).
#modeb(oiliness(var(sushi), num_var(oiliness_val))).
#modeb(seafood(var(sushi))).
#modeb(maki(var(sushi))).
#maxv(1).

#pos(e1, {likes(s1)}, {},          { oiliness(s1,4). seafood(s1). }).
#pos(e2, {likes(s2)}, {},          { oiliness(s2,3). seafood(s2). }).
% too lean
#pos(e3, {},          {likes(s3)}, { oiliness(s3,1). seafood(s3). }).
% too lean
#pos(e4, {},          {likes(s4)}, { oiliness(s4,0). seafood(s4). }).
% oily, not seafood
#pos(e5, {},          {likes(s5)}, { oiliness(s5,4). maki(s5).    }).
% oily, not seafood
#pos(e6, {},          {likes(s6)}, { oiliness(s6,3). maki(s6).    }).
#bias("penalty(1, body(X)) :- in_body(X).").
\end{lstlisting}
\begin{lstlisting}[style=shell]
$ FastLAS --opl examples/ex22_sushi.las
likes(V0) :- oiliness(V0,V_0_oiliness_val), seafood(V0),
        V_0_oiliness_val >= 3, sushi(V0).
\end{lstlisting}
Seafood, oily enough. The threshold \code{3} was never declared: it is the smallest oiliness
among the items the user liked, which is as much as the data supports.


\section{Case studies}
\label{sec:practice}
This section works two applications from the FastLAS literature end to end: the CAVIAR
event-recognition benchmark, and access-control policy learning, the application FastLAS was
originally built for.

\subsection{Event recognition in CAVIAR}
\label{sec:caviar}
CAVIAR is a benchmark of CCTV video in which people are tracked frame by frame. The low-level
activity of each person (\emph{walking}, \emph{running}, \emph{active}, \emph{inactive},
\emph{appear}, \dots) and their pairwise distances are given; the task is to learn definitions of
high-level \emph{composite} events: here, when two people are \emph{meeting}. It is the
flagship scalability benchmark in the FastLAS papers, and it exercises almost everything in this
guide at once: typed variables, \code{const} thresholds, weighted examples,
and computed inclusions and exclusions.

\paragraph{The task.}
Events are modelled in the \emph{Event Calculus}~\cite{eventcalculus}. A \emph{fluent} such as
\code{meeting(P1,P2)}
holds over time; the two predicates we learn, \code{initiatedAt(F,T)} and
\code{terminatedAt(F,T)}, say when a fluent \emph{starts} and \emph{stops} holding, while the
background \code{happensAt(E,T)} records the low-level events. So FastLAS is learning the
\emph{start} and \emph{stop} conditions of a meeting.

\paragraph{The encoding.}
Two excerpts show the structure. First, the background defines the fluent, derives the distance
events \code{close}/\code{further} from raw distances against a threshold, and (using the
computed inclusions and exclusions of Section~\ref{sec:context}) \emph{computes} each example's target
atoms from a \code{goal(holdsAt(...))} annotation:
\begin{lstlisting}[style=fastlas]
fluent(meeting(X,Y)) :- person(X), person(Y).
happensAt(close(Id1,Id2,Th),T)   :- dist(Id1,Id2,T,D), dist(Th),
        D <= Th.
happensAt(further(Id1,Id2,Th),T) :- dist(Id1,Id2,T,D), dist(Th),
        D >  Th.

inclusion(initiatedAt(A,1))  :- not holdsAt(A,1), goal(holdsAt(A,2)).
exclusion(initiatedAt(A,1))  :- fluent(A), not goal(holdsAt(A,2)).
% (symmetric rules give inclusion/exclusion for terminatedAt)

% the const(dist) candidates
dist(24). dist(25). dist(27). dist(34). dist(40).
\end{lstlisting}
Second, the mode declarations. Note \code{const(dist)}, which lets a rule pin a specific distance
threshold drawn from those \code{dist/1} facts, and the trailing \code{
authors use to record each mode's intended recall:
\begin{lstlisting}[style=fastlas]
#modeh(initiatedAt(meeting(var(person), var(person)), var(time))).
#modeh(terminatedAt(meeting(var(person), var(person)), var(time))).

#modeb(happensAt(active(var(person)), var(time))).       %2
#modeb(happensAt(walking(var(person)), var(time))).      %2
#modeb(happensAt(running(var(person)), var(time))).      %2
#modeb(not happensAt(walking(var(person)), var(time))).  %2
#modeb(happensAt(close(var(person),   var(person), const(dist)),
        var(time))).   %10
#modeb(happensAt(further(var(person), var(person), const(dist)),
        var(time))).   %10
% ... and likewise for inactive, abrupt, appear and disappear
\end{lstlisting}
Each example is one frame transition. Its context lists who is present and what they are doing,
the \code{goal(holdsAt(meeting(...),2))} says which meetings should hold next, and the \code{@100}
weight marks it as a (soft) noisy example:
\begin{lstlisting}[style=fastlas]
#pos(p_22732@100, {}, {}, {
  person(id0). person(id1). person(id2). person(id3).
  goal(holdsAt(meeting(id2,id3),2)).
  happensAt(inactive(id0),1). happensAt(running(id2),1).
  happensAt(walking(id3),1).
  % ... plus the raw distances, e.g. dist(id0,id1,1,175). ...
}).
\end{lstlisting}

\paragraph{Running it.}
Read this part as a large-scale worked sketch rather than as a self-contained repository example.

\begin{gotcha}
The data fold is not in the repository accompanying this guide: it ships with the FastLAS source
distribution, as
\code{FastLAS2/data/non_branching_caviar/fastlas_tasks/train_fold_0.las}. Its example ids are
also not unique, so running it as-is stops with \code{Duplicated example ID}. Copy it into the
current directory and make the ids unique first:
\begin{lstlisting}[style=shell]
$ awk '{ if ($0 ~ /^#(pos|neg)\(/) sub(/\(/, "(u" NR "_"); print }' \
      train_fold_0.las > caviar.las
\end{lstlisting}
\end{gotcha}
\begin{lstlisting}[style=shell]
$ FastLAS --opl caviar.las
\end{lstlisting}
This returns a $13$-rule theory. A
representative subset (each rule wraps to fit the page):
\begin{lstlisting}[style=result]
initiatedAt(meeting(V0,V1),V2) :- happensAt(active(V0),V2),
        happensAt(active(V1),V2), happensAt(close(V0,V1,25),V2),
        person(V1), time(V2).
initiatedAt(meeting(V0,V1),V2) :- happensAt(active(V0),V2),
        happensAt(inactive(V1),V2), happensAt(close(V1,V0,24),V2),
        person(V0), person(V1), time(V2).
terminatedAt(meeting(V0,V1),V2) :- happensAt(running(V1),V2),
        person(V1), time(V2).
terminatedAt(meeting(V0,V1),V2) :- happensAt(walking(V0),V2),
        happensAt(inactive(V1),V2), person(V1), time(V2).
\end{lstlisting}
In English: a meeting \emph{starts} when both people are \code{active} and \code{close} within
distance~25, or when one is \code{active} beside an \code{inactive} companion within~24; it
\emph{ends} when one of them starts \code{running}, or walks off while the other is
\code{inactive}. The \code{25} and the \code{24} are \code{const(dist)} values
FastLAS selected from the \code{dist/1} pool: the payoff of the \code{const} mechanism from
Section~\ref{sec:modes}. The thirteen-rule size is stable from run to run; \emph{which} of
several equal-cost rules survives is not (the \code{const(dist)} thresholds in particular move
between runs), so treat the listing above as one run's output rather than as the answer. This is
the scale, tens of thousands of examples, that motivates FastLAS in the first place.

\subsection{Access-control policy learning}
\label{sec:policy}
Example~\ref{ex:policy} showed a policy learned from a toy log. The experiments of
\citeN{fastlas} do the same at full scale, on access logs generated from the \emph{Project
Management} case study of \citeN{xustoller}: requests by users (\emph{subjects}) to act on
\emph{resources}, each granted or denied by a ground-truth policy. The task files ship with the
FastLAS source distribution, under \code{FastLAS1/data/policy_learning/files/}; this subsection
reproduces the length-scored (\code{sf_len}) experiment that learns \code{accept}.

\paragraph{The task.}
Each logged request is one example. Its context lists the request's attributes: who is asking
(\code{subject_role}, \code{subject_projects}, \dots), for what
(\code{resource_type}, \code{resource_project_id}, \dots) and to do what
(\code{action_action_type}); thirteen attributes in all. A granted request includes
\code{accept} and a denied one excludes it, exactly as in Example~\ref{ex:policy}; the first of
the ten cross-validation folds has 409 grants and 40 denials. The learned rules say when to
accept, and a request matched by no rule is denied.

\paragraph{The encoding.}
The mode declarations give each attribute one \code{const} body mode, and the \code{#bias} lines
implement the paper's $S_{\mathit{len}}$ objective, one unit per literal, exactly as in
Section~\ref{sec:bias}:
\begin{lstlisting}[style=fastlas]
#modeh(accept).
#modeb(1, subject_role(const(subject_role))).
#modeb(1, resource_type(const(resource_type))).
#modeb(1, action_action_type(const(action_action_type))).
% ... ten more attribute modes ...
#maxv(1).
#bias("penalty(1, head(X)) :- in_head(X).").
#bias("penalty(1, body(X)) :- in_body(X).").
\end{lstlisting}
One example of each kind, contexts abbreviated:
\begin{lstlisting}[style=fastlas]
#pos(eg(id0), {accept}, {}, {
  subject_role(employee).      resource_type(task).
  action_action_type(read).    % ... eight more attributes ...
}).
#pos(eg(id409), {}, {accept}, {
  subject_role(projectleader). resource_type(task).
  action_action_type(approve). % ...
}).
\end{lstlisting}
\begin{gotcha}
The files are in FastLAS~1 syntax, and declare their constant pools with a directive that
FastLAS~2 no longer accepts: \code{#constant(subject_role, employee).} and so on. In FastLAS~2 a
pool is ordinary background facts (Section~\ref{sec:modes}), but do \emph{not} port
\code{#constant(t, v).} to \code{t(v).} when \code{t} is also a context predicate, as it is
here. The pool facts would then hold in \emph{every} example, so every candidate body would be
true everywhere, no rule could avoid the denials, and the task would come back
\code{UNSATISFIABLE}. Give each pool a fresh predicate instead:
\begin{lstlisting}[style=shell]
$ sed -E \
    -e 's/^#constant\(([a-z_]+), *([A-Za-z0-9_]+)\)\./\1_t(\2)./' \
    -e 's/const\(([a-z_]+)\)/const(\1_t)/g' \
    PM_full_fold_0_accept_sf_len_train.las > pm_fold0_len.las
\end{lstlisting}
\end{gotcha}
\noindent The converted fold is \code{examples/pm_fold0_len.las} in the repository accompanying
this guide.

\paragraph{Running it.}
\begin{lstlisting}[style=shell]
$ FastLAS --opl examples/pm_fold0_len.las
\end{lstlisting}
This takes about twenty seconds on the machine of Section~\ref{sec:perf} and returns a theory
of around 55 rules (53 to 59 across folds and runs). Some are genuinely general, others
memorise a single user or resource:
\begin{lstlisting}[style=result]
accept :- action_action_type(setschedule), subject_role(planner).
accept :- resource_is_properietary(true), subject_role(employee).
accept :- subject_user_id(pjl3__0).
accept :- resource_dept_id(dept__4).
\end{lstlisting}
The distribution also ships each held-out fold as a clingo program that tries a theory on the
unseen requests and classifies each as \code{tp}, \code{fp}, \code{tn} or \code{fn}. Save the
learned rules to \code{hyp.lp} and count:
\begin{lstlisting}[style=shell]
$ clingo --enum-mode=brave --quiet=1 -n 0 hyp.lp \
      PM_full_fold_0_accept_test.las
\end{lstlisting}
Repeating train-and-count over the ten folds gives precision $0.954$, recall $0.905$ and $F_1$
$0.929$; the published results for this experiment (Table~4 of \citeN{fastlas}) are $0.951$,
$0.905$ and $0.928$. The port reproduces the paper.

The memorised rules are no accident. Every example above is hard, so each training grant that
no general rule explains must be bought a rule of its own. The paper's objective, however, is
$S_{\mathit{len}}$ \emph{plus} the penalty of uncovered examples, and weighted examples
(Example~\ref{ex:weights}) express exactly that: \code{examples/pm_fold0_len_pen.las} differs
only in an \code{@1} on every example id, making an outlier cost one unit to leave uncovered
but at least two to memorise. The same command then returns around nine rules:
\begin{lstlisting}[style=result]
accept :- action_action_type(read).
accept :- subject_role(employee).
accept :- resource_type(budget), subject_role(manager).
accept :- action_action_type(setcost), subject_role(accountant).
\end{lstlisting}
a policy readable at a glance (anyone may read, employees may act, managers handle budgets,
accountants set costs, \dots), at precision $0.950$, recall $1.000$ and $F_1$ $0.974$ over the
ten folds. As with CAVIAR, \emph{which} equal-cost rules survive changes from run to run; the
sizes and the scores are what stay stable.



\section{Running FastLAS and reading its output}
\label{sec:running}
\subsection{Command-line flags}
One of \code{--opl} or \code{--nopl} must always appear; the rest are occasional. These are the
ones used in this manual.
\begin{center}
\begin{tabular}{@{}ll@{}}
\toprule
\texttt{-{}-opl} / \texttt{-{}-nopl} & choose the algorithm (one is required) \\
\texttt{-{}-version}                 & print the version \\
\texttt{-{}-help}                    & list all options \\
\texttt{-{}-debug}                   & verbose trace of the algorithm's phases \\
\texttt{-{}-force-safety}            & enforce the safety constraint on learned rules \\
\texttt{-{}-threads N}               & parallelism; the default is \code{8} \\
\texttt{-{}-timeout T}               & time limit for the final solving stage \\
\texttt{-{}-score-only}              & print only the score of the solution \\
\texttt{-{}-space-size}              & prepend the size of the kept candidate space \\
\texttt{-{}-output-solve-program}    & print the final ASP program and stop \\
\texttt{-{}-max-conditions N}        & bounded numeric variables per rule (Section~\ref{sec:modes}) \\
\texttt{-{}-write-cache F} / \texttt{-{}-read-cache F} & save/reuse the computed state (Section~\ref{sec:incremental}) \\
\bottomrule
\end{tabular}
\end{center}

\subsection{Interpreting the output}
FastLAS is terse: it prints the hypothesis and nothing else. Five things can come back, and it is
worth knowing what each one means before you start changing your task.
\begin{itemize}[nosep,leftmargin=1.4em]
  \item \textbf{One or more rules}: the learned hypothesis. Learned variables are named
        \code{V0}, \code{V1}, \dots, and type atoms may be appended for safety. The rule is
        stable but \emph{the order of its body literals is not}: repeated runs of the same task
        may print the same conjunction in a different order, and may name the variables
        differently. Compare hypotheses as sets of literals, not as strings.
  \item \textbf{(blank)}: the empty hypothesis is optimal, e.g.\ all examples are weighted and
        cheaper to leave uncovered. Note that when the background already entails an example,
        \code{--nopl} prints nothing but \code{--opl} still emits a redundant rule.
  \item \texttt{UNSATISFIABLE}: no hypothesis in the space explains the (hard) examples.
        Common causes: a needed rule is outside the modes; \code{#maxv} too low; the
        task is non-observational but you used \texttt{-{}-opl}; the scenario was left out of the
        context. A task containing a \code{#neg} also needs \code{--nopl}
        (Section~\ref{sec:context}). If \emph{every} task reports this, including a known-good
        one such as \code{examples/ex01_cycle.las}, check the installation first
        (Section~\ref{sec:whatyouneed}).
  \item \textbf{Two hypotheses, each with a JSON summary}: your task contains a
        \code{#predict}; the output is a satisfying/not-satisfying pair (Section~\ref{sec:predict}).
  \item \texttt{Unknown token: '\#'} or \texttt{syntax error \dots}: a parse error; check for
        unsupported directives such as \code{#constant}, or a missing
        context slot on \code{#neg}.
\end{itemize}
\subsection{Seeing the solve program: \texttt{-{}-output-solve-program}}
\code{--output-solve-program} runs the FastLAS algorithm up to the final search, then prints the
assembled ASP optimisation program to stdout and stops (no solving). It is the best window into a
task that is unexpectedly \code{UNSATISFIABLE} or slow. Read it, or run it in Clingo yourself:
\begin{lstlisting}[style=shell]
$ FastLAS --opl --output-solve-program examples/ex01_cycle.las
\end{lstlisting}
\begin{lstlisting}[style=result]
% d1
% d1 : d1
disj(0) :- in_h(0).
n_cov(d1) :- not disj(0).
n_cov(d1) :- disj(1).
n_cov(d1) :- #true, n_cov(d1).
:- n_cov(d1).
% d2
% d2 : d2
n_cov(d2) :- disj(1).
n_cov(d2) :- #true, n_cov(d2).
:- n_cov(d2).
0 {in_h(0)} 1.
rule_score(0, 2).
% ... plus the :~ weak constraints, #show, and the lua #script block
\end{lstlisting}
Each \code{in_h(i)} is a candidate rule the search may switch on; \code{disj(i)} records that the
\code{i}th of an example's coverage conditions is met; \code{n_cov(eg)} marks an example the
current hypothesis fails to cover, and the \code{:- n_cov(eg).} constraints force coverage. The
\code{
and the index \code{i} is assigned afresh on each run, so a rerun may print \code{in_h(1)} where
this transcript shows \code{in_h(0)}. One caveat: on a \code{#predict} task only the
prediction-\emph{satisfying} program is emitted.

\paragraph{Diagnosing an \texttt{UNSATISFIABLE} task.}
The flag earns its keep when a task fails and you cannot see why. Consider this one, which asks for
a relation between two nodes but permits only one variable per rule:
\begin{lstlisting}[style=fastlas]
% Deliberately UNSATISFIABLE: the target needs two variables, but
% #maxv(1) allows only one, so no candidate rule can cover p1.
% Section 8.3 diagnoses it.
node(a). node(b). edge(a,b).
#modeh(rel(var(node), var(node))).
#modeb(edge(var(node), var(node))).
#maxv(1).
#pos(p1, {rel(a,b)}, {}, {}).
#bias("penalty(1, body(X)) :- in_body(X).").
\end{lstlisting}
\begin{lstlisting}[style=shell]
$ FastLAS --opl examples/ex30_unsat_diagnosis.las
UNSATISFIABLE
\end{lstlisting}
Nothing there says what went wrong. The solve program does:
\begin{lstlisting}[style=shell]
$ FastLAS --opl --output-solve-program examples/ex30_unsat_diagnosis.las
% p1
% p1 : p1
n_cov(p1) :- not disj(0).
n_cov(p1) :- disj(0).
n_cov(p1) :- #true, n_cov(p1).
:- n_cov(p1).
\end{lstlisting}
Skip the comment and the vacuous self-loop and read the remaining three lines together. The first
two say that \code{p1} is uncovered when \code{disj(0)}
fails \emph{and} when it holds, so \code{n_cov(p1)} is derivable no matter what the search
chooses; the third then forbids exactly that. The program is contradictory on its own, which is
the signature of an example no candidate rule can satisfy. \code{--space-size} confirms it from
the other direction:
\begin{lstlisting}[style=shell]
$ FastLAS --opl --space-size examples/ex30_unsat_diagnosis.las
% SPACE SIZE: 0
UNSATISFIABLE
\end{lstlisting}
There are no candidates at all, because \code{rel(var(node), var(node))} needs two variables and
\code{#maxv(1)} allows one. Raise it to \code{#maxv(2)} and the task is solved. The habit worth
forming: when a task is \code{UNSATISFIABLE}, look at \code{--space-size} first, since a space of
\code{0} means the fault is in the mode declarations rather than in the examples.


\section{FastLAS vs.\ ILASP: a porting cheat-sheet}
\label{sec:ilasp}
If you are moving a task between the two systems, most of it carries over unchanged: the
example blocks (\code{#pos}/\code{#neg} with \code{@penalty}), \code{#maxv}, and the
\code{#modeh}/\code{#modeb} predicate skeletons are written the same way. This section covers
what does not: whether a given task is worth porting at all, what the line-by-line changes are,
how the directives correspond, and what to do when a port fails.

\subsection{Deciding whether the task ports at all}
\label{sec:portable}
Before touching the syntax, ask what shape the answer has. If the thing you want to learn is a
set of \emph{definite conclusions}, a definition or a constraint, FastLAS will do it and will
scale further. If what you want is a \emph{space} of possibilities or an \emph{ordering} over
them, the task is ILASP's, and no amount of rewriting will move it.
\begin{center}\small
\begin{tabular}{@{}p{0.44\linewidth}p{0.5\linewidth}@{}}
\toprule
\textbf{If the target is\dots} & \textbf{then} \\
\midrule
a definition, \code{p :- q, not r.} & port it; this is FastLAS's home ground \\
a constraint, ``never both'' & port it as a \code{violated} head (Section~\ref{sec:limits}) \\
a numeric threshold & port it, and use \code{num_var} (Section~\ref{sec:modes}) \\
a domain-specific notion of ``best'' & port it \emph{to} FastLAS; ILASP cannot express it \\
a choice, ``either heads or tails'' & stay in ILASP (Exercise 22) \\
a preference or ranking & stay in ILASP (Section~\ref{sec:preferences}) \\
a recursive definition & stay in ILASP, or move the recursion into the background \\
\bottomrule
\end{tabular}
\end{center}

\subsection{A task ported line by line}
Here are the opening lines of the same learning problem in both systems, recognising an animal
that flies. Only the head of each file is shown, enough to see the four differences; a task that
actually learns the flightless exception needs the counter-example of
Example~\ref{ex:flies-exc} as well. On the left, ILASP; on the right, FastLAS.
\begin{lstlisting}[style=fastlas,breaklines=false]
% ILASP                          | % FastLAS
#modeh(flies(var(animal))).      | #modeh(flies(var(animal))).
#modeb(winged(var(animal))).     | #modeb(winged(var(animal))).
#modeb(flightless(var(animal))). | #modeb(not flightless(var(animal))).
#maxv(1).                        | #maxv(1).
#constant(animal, eagle).        | animal(eagle).      % a fact
#pos(p1, {flies(eagle)}, {}).    | #pos(p1, {flies(eagle)}, {},
                                 |      { winged(eagle). }).
\end{lstlisting}
Three changed lines, and a fourth difference that does not show up in a listing this small:
\begin{itemize}[nosep,leftmargin=1.4em]
  \item \textbf{Negation is explicit.} ILASP generates the negated version of each \code{#modeb}
        for you (unless you write \code{(positive)}); FastLAS does not, so declare
        \code{#modeb(not p(...))} for every negated literal you want available.
  \item \textbf{Constants are facts.} \code{#constant(t, v).} is not a FastLAS directive and will
        stop the parser with \texttt{Unknown token: '\#'}. Write \code{t(v).} in the background or,
        better, in the context of the examples that need it.
  \item \textbf{Examples carry their scenario.} ILASP examples often lean on one global background;
        FastLAS expects the per-example data in the fourth slot. This is the single most common
        cause of a ported task returning \code{UNSATISFIABLE} while nothing looks wrong.
  \item \textbf{\texttt{\#bias} changes meaning.} In ILASP it prunes the search space; in FastLAS it
        assigns cost. A ported \code{#bias} is almost never still correct, and a silently wrong one
        is worse than a parse error.
\end{itemize}

\subsection{How the directives correspond}
With the shape of a port settled, this is the reference table: what each system calls a given
feature, and where FastLAS simply has no counterpart.
\begin{center}\small\raggedright
\begin{tabular}{@{}p{0.235\linewidth}p{0.37\linewidth}p{0.335\linewidth}@{}}
\toprule
\textbf{Concept} & \textbf{ILASP} & \textbf{FastLAS 2.2.0} \\
\midrule
Head declarations      & \code{#modeh}, \code{#modeha} (choice) & \code{#modeh} only \\
Body declarations      & \code{#modeb}, auto-NAF                & \code{#modeb}; explicit \code{not} decls \\
Condition/aggregates  & \code{#modec}                          & not supported \\
Weak-constraint bias   & \code{#modeo},\code{#weight},\code{#maxp} & not supported \\
Mode options           & \code{(positive)},\code{(symmetric)}\dots & not supported \\
Constants              & \code{#constant(t,v)}                  & \code{t(v)} facts in contexts \\
Max variables          & \code{#maxv}                           & \code{#maxv} (same) \\
Hypothesis-size cap    & \code{#max_penalty} (default 15)       & \code{#max_penalty}; accepted, no default cap \\
\code{#bias("...")}    & pruning constraints                    & \textbf{scoring} (\code{penalty}/\code{in_body}) \\
Objective              & length + penalties (fixed)             & {\raggedright\textbf{user-defined}; no objective unless you write one\par} \tabularnewline
Ordering/preferences   & \code{#brave_ordering}, \code{#cautious_ordering} & not supported \\
Recursion, choice      & yes                  & no \\
CLI                    & \code{ILASP --version=N task}          & \code{FastLAS --opl/--nopl task} \\
\bottomrule
\end{tabular}
\end{center}

\subsection{What to do when a port fails}
Most failures announce themselves as a parse error or an unexpected \code{UNSATISFIABLE}. The
table maps the message you see to the thing that usually caused it.
\begin{center}\small
\begin{tabular}{@{}p{0.36\linewidth}p{0.58\linewidth}@{}}
\toprule
\textbf{Symptom} & \textbf{Usual cause} \\
\midrule
\texttt{Unknown token: '\#'}   & \code{#constant}, \code{#modeo}, \code{#modec}, or a Clingo aggregate \\
\texttt{unexpected T\_BASIC\_SYMBOL} & \code{#modeha}, i.e.\ an ILASP choice head \\
\texttt{syntax error, unexpected T\_COLON} & a conditional literal \code{p : q}, or a weak constraint \code{:~} \\
\code{UNSATISFIABLE}           & the scenario is not in the context slot; or a negated body mode was never declared; or \code{#maxv} is too low \\
a rule that is too general     & no example rules out the shorter candidate (Example~\ref{ex:vertexcover}) \\
the wrong rule of several ties & the objective, not the data: add a \code{#bias} (Section~\ref{sec:bias}) \\
\bottomrule
\end{tabular}
\end{center}
\noindent When none of these applies, \code{--output-solve-program} (Section~\ref{sec:running})
prints the exact ASP that FastLAS is about to solve, which is usually enough to see which example
is doing the damage.


\section{Quick reference}
\label{sec:files}
The user-facing directives of FastLAS 2.2.0:
\begin{center}\small
\begin{tabular}{@{}ll@{}}
\toprule
\code{#modeh(A).}                  & head atom \code{A} allowed in learned rule heads \\
\code{#modeh(false).}              & learn a constraint-style verifier with head \code{false} \\
\code{#modeb(A).} / \code{#modeb(not A).} & body literal allowed (optionally negated) \\
\code{#modeb(N, A).}               & \dots with recall bound \code{N} \\
\code{#maxv(N).}                   & max distinct variables per rule \\
\code{#max_penalty(N).}            & cap on hypothesis cost \\
\code{#bias("...").}               & scoring program (costs); see Section~\ref{sec:bias} \\
\code{#final_bias("...").}         & second-stage scoring \\
\code{#pos(id,{I},{E},{C}).}       & positive example (3-slot form also allowed) \\
\code{#neg(id,{I},{E},{C}).}       & negative example (context slot mandatory) \\
\code{#pos(id@W, ...).}            & weighted (noisy) example, penalty \code{W} \\
\code{#predict(id,{I},{E},{C}).}   & prediction query (advanced) \\
\bottomrule
\end{tabular}
\end{center}
Inside modes: \code{var(t)} typed variable, \code{const(t)} typed constant, \code{num_var(t)}
numeric variable that learns \code{>=}/\code{<=} bounds (Section~\ref{sec:modes}). Comments: \code{

\paragraph{The example files.}
Every numbered example in this guide, except where the text explicitly points to an external
artifact, and every exercise solution of \ref{app:solutions}, is a task file in the accompanying
repository:
\begin{center}
\url{https://github.com/dasaro/fastlas_manual}
\end{center}
The examples are under \texttt{examples/} and the solutions under
\texttt{examples/solutions/}. Clone it and each one runs directly against FastLAS 2.2.0, e.g.
\begin{lstlisting}[style=shell]
$ git clone https://github.com/dasaro/fastlas_manual.git
$ cd fastlas_manual
$ FastLAS --opl examples/ex03_flies_exception.las
\end{lstlisting}


\section{Exercises}
\label{sec:exercises}
The exercises below run through the whole guide, from a first task to the case-study features.
Every exercise has a solution: selected ones are worked in \ref{app:solutions}, and the rest are
tested task files in \code{examples/solutions/} of the repository (Section~\ref{sec:files}),
named where the appendix points to them. Every reference solution was checked against FastLAS
2.2.0. Try yours, then compare the learned hypothesis, not just the file. Unless stated
otherwise, run with \code{--opl}.

\subsection*{Warm-up (the first task and the anatomy of a file)}

\paragraph{Exercise 1 (a two-hop rule).}
A family is given by \code{parent/2} facts: \code{parent(ann,bob)}, \code{parent(bob,carl)},
\code{parent(carl,dee)}. Declare modes (head \code{grandparent}, body \code{parent},
\code{#maxv(3)}) and give one example whose \emph{include} set holds the true grandparent pairs and
whose \emph{exclude} set holds the near-misses (parents and great-grandparents), so that FastLAS is
forced to learn the two-\code{parent} join rather than a shorter over-general rule.
Starter/solution: \code{examples/solutions/sol01_grandparent.las}.

\paragraph{Exercise 2 (fix a broken task).}
The task below returns \code{UNSATISFIABLE}: no rule over the given modes can make winged
\code{ostrich} \emph{not} fly while eagles and sparrows do. Add exactly one line to fix it.
\begin{lstlisting}[style=fastlas]
animal(eagle). animal(ostrich). animal(sparrow).
#modeh(flies(var(animal))).
#modeb(winged(var(animal))).
#maxv(1).
#pos(p1, {flies(eagle)},   {}, { winged(eagle).   }).
#pos(p2, {flies(sparrow)}, {}, { winged(sparrow). }).
#pos(p3, {}, {flies(ostrich)},
        { winged(ostrich). flightless(ostrich). }).
\end{lstlisting}
Solution: \code{examples/solutions/sol02_exception.las}.

\paragraph{Exercise 3 (pick the constant).}
With \code{#modeb(chosen(const(colour)))} and colours \code{red}, \code{blue}, \code{green}, learn
which single colour makes \code{sel} true, given one accepted context \code{chosen(blue)} and two
rejected ones (\code{chosen(red)}, \code{chosen(green)}).
Solution: \code{examples/solutions/sol03_const.las}.

\paragraph{Exercise 4 (steer the scoring).}
Both \code{sel :- a.} and \code{sel :- b.} explain the data below equally well, and plain length
returns \code{sel :- a.}. Change \emph{only} the \code{#bias} so FastLAS returns \code{sel :- b.}
instead.
\begin{lstlisting}[style=fastlas]
#modeh(sel).  #modeb(a).  #modeb(b).
#pos(p1, {sel}, {},    { a. b. }).
#pos(p2, {},    {sel}, { c. }).
#bias("penalty(1, body(a)) :- in_body(a).").
#bias("penalty(1, body(b)) :- in_body(b).").
\end{lstlisting}
Solution: \code{examples/solutions/sol04_bias.las}.

\subsection*{Intermediate (modes, algorithms and effective programs)}

\paragraph{Exercise 5 (a learned threshold).}
Using \code{num_var}, learn \code{adult} from \code{age/2} observations: positives at ages
\code{20} and \code{40}, negatives at \code{10} and \code{17}. Keep \code{#maxv(1)}. Which bound does
FastLAS synthesise, and why that number rather than, say, \code{>= 18}?
Solution: \code{examples/solutions/sol05_numvar.las}.

\paragraph{Exercise 6 (two thresholds in one rule).}
Cars are \code{ok} exactly when \emph{both} a speed band and a weight band hold, and the data is
arranged so neither attribute alone separates the classes. The default run prints
\code{UNSATISFIABLE}. Which flag lets FastLAS bound two numeric variables in a single rule, and what
rule results? Solution: \code{examples/solutions/sol06_numvar_multi.las}.

\paragraph{Exercise 7 (choose the algorithm).}
In the task below \code{permitted} never appears in any example; it only influences the observed
\code{fault} through the background. Explain why \code{--opl} returns \code{UNSATISFIABLE}, and run
it so it succeeds.
\begin{lstlisting}[style=fastlas]
fault :- did(A), not permitted(A), act(A).
#modeh(permitted(var(act))).
#modeb(authorised(var(act))).
#maxv(1).
#pos(e1, {},      {fault}, { act(a1). did(a1). authorised(a1). }).
#pos(e2, {fault}, {},      { act(a2). did(a2). }).
\end{lstlisting}
Solution: \code{examples/solutions/sol07_nopl.las}.

\paragraph{Exercise 8 (translate a Clingo idiom).}
FastLAS rejects the aggregate below with \code{Unknown token: '#'}. Rewrite it as an auxiliary rule
in FastLAS's dialect, then complete a task that learns \code{safe(X) :- not attacked(X)}.
\begin{lstlisting}[style=fastlas]
attacked(X) :- #count{ Y : att(Y,X) } >= 1.   % illegal in FastLAS
\end{lstlisting}
Solution: \code{examples/solutions/sol08_dialect.las}.

\paragraph{Exercise 9 (ask a prediction).}
Take the cycle/rain task (\code{cycle :- not rain.} is learnable) and add a \code{#predict} asking
whether ``\code{cycle} holds on a dry day'' (\code{cycle} included, empty context). From the
two-hypothesis output, is that conclusion \emph{entailed}? Solution:
\code{examples/solutions/sol09_predict.las}.

\subsection*{Advanced (scoring, limits, streams and the case studies)}

\paragraph{Exercise 10 (reframe a recursive target).}
Take the argumentation framework with arguments \code{a}, \code{b}, \code{c} and attacks
\code{att(a,b)} and \code{att(b,c)} (Section~\ref{sec:limits} explains the setting). Build a
verifier task for it: each context carries the framework and one labelling, conflict-free
labellings (such as \code{a} and \code{c} in, \code{b} out) \emph{exclude} \code{violated}, and
labellings with an attack between two \code{in} arguments \emph{include} it. Check that FastLAS
learns the same conflict-freeness rule as Example~\ref{ex:verifier}, and say why learning
\code{in/1} directly would not have worked. Solution:
\code{examples/solutions/sol10_verifier.las}.

\paragraph{Exercise 11 (score a feature once).}
The only rule that fits the data is \code{p :- not a, not b} (two negations). Using \code{#final_bias}
with an \code{intermediate/1} feature, charge a \emph{single} unit for using negation-as-failure,
regardless of how many negated literals appear. Confirm with \code{--score-only} that the winning
rule scores \code{1}, not \code{2}. Solution: \code{examples/solutions/sol11_finalbias.las}.

\paragraph{Exercise 12 (learn from a stream).}
Window~1 holds one example forcing \code{p.}; window~2 adds a contradicting example. With
\code{--write-cache}/\code{--read-cache}, show that the incremental result after window~2 equals
the batch result on both windows, and differs from window~1's answer. Solution files:
\code{examples/solutions/sol12_window1.las}, \code{sol12_window2.las}, \code{sol12_batch.las}.

\paragraph{Exercise 13 (change the objective, change the policy).}
The access log of Example~\ref{ex:policy} yields \code{accept :- subject_role(manager).} under
plain length. Add one domain
\code{#bias} that charges for role conditions so FastLAS instead prefers the clearance-based policy
\code{accept :- subject_clearance(high).} Solution: \code{examples/solutions/sol13_policy.las}.

\subsection*{From the classic problems (Section \ref{sec:classics})}

\paragraph{Exercise 14 (the other half of the cover).}
Example~\ref{ex:vertexcover} learns \code{covered(X,Y) :- edge(X,Y), in(X).} The Clingo program
also has the symmetric rule, covering an edge by its \emph{second} endpoint. Change the examples
so that FastLAS learns that one instead, and keep the task down to the four examples used before.
Solution: \code{examples/solutions/sol14_cover_other.las}.

\paragraph{Exercise 15 (which streets are dangerous).}
The path problem of the course notes labels each street with a danger score,
\code{dangerous(a,b,4)}, \code{dangerous(a,c,1)}, \code{dangerous(b,c,3)},
\code{dangerous(c,d,1)}, \code{dangerous(b,d,1)}. Suppose the two streets scoring 3 or more,
\code{a,b} and \code{b,c}, are considered \code{risky} and the three scoring 1 are not. Use \code{num_var} to learn the threshold rather than writing
it, and say why the bound comes out where it does.
Solution: \code{examples/solutions/sol15_dangerous.las}.

\subsection*{With pen and paper}
These three want no computer at all, though you can check each answer with one command afterwards.

\paragraph{Exercise 16 (enumerate a search space).}
Take this mode bias, which is propositional, so no variables are involved:
\begin{lstlisting}[style=fastlas]
#modeh(p).
#modeb(a).
#modeb(b).
\end{lstlisting}
Write out, by hand, every rule that FastLAS may consider. Then answer a second question: ILASP's
\code{-s} flag reports \emph{nine} rules with head \code{p} for the same three lines. Why more,
and what would you have to add to the FastLAS bias to obtain the same nine?

\paragraph{Exercise 17 (compute a score).}
The task below learns \code{q :- r.} Before running anything, work out what
\code{--score-only} will print.
\begin{lstlisting}[style=fastlas]
#modeh(q).
#modeb(r).
#modeb(s).
#pos(e1, {q}, {}, { r. s. }).
#pos(e2, {}, {q}, { s. }).
#bias("penalty(2, head)    :- in_head(X).").
#bias("penalty(3, body(X)) :- in_body(X).").
\end{lstlisting}

\paragraph{Exercise 18 (why is nothing the best answer?).}
Take Example~\ref{ex:weights} and swap its two penalties, so that \code{e1} costs~1 to abandon and
\code{e2} costs~3. FastLAS then returns the empty hypothesis. Explain, in terms of the objective
of Section~\ref{sec:whatitcomputes}, why learning \emph{nothing} is optimal here, and say what the
score is.

\subsection*{At the keyboard}

\paragraph{Exercise 19 (negative examples).}
A car is roadworthy unless it has failed its inspection. You are given
\code{car(c1). car(c2). failed(c2).}, a head mode \code{roadworthy(var(car))}, body modes
\code{car(var(car))} and \code{not failed(var(car))}, \code{#maxv(1)}, and the single positive
example \code{#pos(p1, {roadworthy(c1)}, {}, {}).} As it stands, FastLAS answers
\code{roadworthy(V0) :- car(V0).}, which would make the failed car roadworthy too. Add one
negative example that rules this out, and say which algorithm flag you now need.
Solution: \code{examples/solutions/sol17_neg_roadworthy.las}.

\paragraph{Exercise 20 (what a penalty buys).}
Take the two contradictory examples of Example~\ref{ex:weights}, with \code{e2} fixed at
penalty~1. Raise the penalty on \code{e1} one step at a time and find the smallest value at which
FastLAS stops returning the empty hypothesis and commits to \code{p.} Explain the number you
find by adding up the two costs it is comparing.
Solution: \code{examples/solutions/sol18_weights.las}.

\paragraph{Exercise 21 (build a task only \texttt{-{}-nopl} can solve).}
The exercises so far have handed you the task. Now write one. Construct, from scratch, a FastLAS
task that returns a hypothesis under \code{--nopl} and \code{UNSATISFIABLE} under \code{--opl},
using no more than a handful of lines. State which property of your task is responsible, and check
both runs. There is more than one way to do it.
Solution: \code{examples/ex31_nopl_only.las}.

\paragraph{Exercise 22 (build a task only ILASP can solve).}
Harder, and worth the effort. Construct a learning task that ILASP solves but FastLAS cannot,
where the obstacle is the \emph{shape of the hypothesis} rather than a syntactic detail of the
input. Write the task in both dialects, run \code{ILASP --version=4} on one and FastLAS on the
other, and explain in a sentence what it is about the rules ILASP returns that FastLAS will never
produce. Section~\ref{sec:portable} says where to look.
Solution: \code{examples/ex32_coins_ilasp.las} with
\code{examples/solutions/sol16_coins.las}.

\paragraph{Exercise 23 (a task FastLAS cannot do).}
Three coins are flipped and the outcome of each is observed, as in the introductory ILASP task of
the course notes, one flip being
\code{#pos(f1, {heads(c1), tails(c2), heads(c3)}, {tails(c1), heads(c2), tails(c3)}, {}).}
ILASP learns the pair \code{heads(V) :- not tails(V), coin(V).} and
\code{tails(V) :- not heads(V), coin(V).} Pose the same task to FastLAS, with both predicates as
heads and their negations as body literals. What comes back, and what is it about this particular
hypothesis that FastLAS will not produce?
Solution: \code{examples/solutions/sol16_coins.las}.



\appendix
\section{Solutions to the exercises}
\label{app:solutions}
Every hypothesis below is the verbatim output of FastLAS 2.2.0 on the corresponding file in
\code{examples/solutions/}. FastLAS appends the type atoms (\code{person(V0)}, \code{car(V0)},
\dots) that its \code{var(t)} placeholders stand for; they are part of the learned rule.

\noindent This appendix is structured in three layers: selected worked solutions from the earlier
tutorial sections, a complete worked set for the \emph{With pen and paper} exercises, and a
complete worked set for the \emph{At the keyboard} exercises.

\subsection*{Selected earlier solutions in task-file form}
The remaining early solutions are mechanical, and the tested task files carry them: each of those
exercises names its file, and all of them live in \code{examples/solutions/}.

\paragraph{Solution 5.}
\code{FastLAS --opl examples/solutions/sol05_numvar.las} learns a one-sided lower bound:
\begin{lstlisting}[style=result]
adult(V0) :- age(V0,V_0_age_val), V_0_age_val >= 20, person(V0).
\end{lstlisting}
FastLAS synthesises \code{>= 20}, the smallest \emph{observed} positive value, not the ``textbook''
\code{18}: it can only propose bounds at values that appear in the data. Because every negative sits
below the positives, the bound is one-sided.

\paragraph{Solution 6.}
The default caps a rule at one bounded numeric variable, so no single rule can bound both speed and
weight and the task is \code{UNSATISFIABLE}. Raise the limit with \code{--max-conditions 2}:
\begin{lstlisting}[style=shell]
$ FastLAS --opl --max-conditions 2 \
      examples/solutions/sol06_numvar_multi.las
\end{lstlisting}
\begin{lstlisting}[style=result]
ok(V0) :- speed(V0,V_0_speed_val), weight(V0,V_0_weight_val),
          V_0_weight_val >= 1400, V_0_speed_val >= 70,
          V_0_weight_val <= 1500, V_0_speed_val <= 80, car(V0).
\end{lstlisting}
(\code{--num-var-count} would \emph{not} help here: that flag adds numeric slots per type, while
the gate on two \emph{different} bounded quantities in one rule is \code{--max-conditions}.)

\paragraph{Solution 7.}
\code{permitted} is \emph{non-observational}: it appears in no example, so the OPL algorithm has
nothing to generalise from and returns \code{UNSATISFIABLE}. FastNonOPL abduces it:
\begin{lstlisting}[style=shell]
$ FastLAS --nopl examples/solutions/sol07_nopl.las
\end{lstlisting}
\begin{lstlisting}[style=result]
permitted(V0) :- authorised(V0), act(V0).
\end{lstlisting}

\paragraph{Solution 9.}
The prediction is \textbf{entailed}. FastLAS prints two hypotheses: the one \emph{satisfying} the
query is the normal optimum \code{cycle :- not rain.}, while the one \emph{not satisfying} it comes
back \code{UNSATISFIABLE}: there is no optimal theory that refutes the query, so it must hold.
\begin{lstlisting}[style=result]
% Optimal hypothesis satisfying the prediction:
cycle :- not rain.
% Optimal hypothesis not satisfying the prediction:
UNSATISFIABLE
\end{lstlisting}

\paragraph{Solution 11.}
The \code{#bias} builds one aggregate feature, \code{intermediate(naf)}, that is on as soon as the
rule uses any negation; \code{#final_bias} then charges for the feature, not per literal:
\begin{lstlisting}[style=fastlas]
#bias("intermediate(naf) :- in_body(neg(X)).").
#final_bias("penalty(1, naf) :- intermediate(naf).").
\end{lstlisting}
\code{FastLAS --opl examples/solutions/sol11_finalbias.las} learns \code{p :- not b, not a.}, and
\code{FastLAS --opl --score-only examples/solutions/sol11_finalbias.las} prints \code{1}: one
charge for the whole
rule, though it has two negations.

\paragraph{Solution 15.}
\code{FastLAS --opl examples/solutions/sol15_dangerous.las} learns a lower bound on the danger score:
\begin{lstlisting}[style=result]
risky(V0,V1) :- dangerous(V0,V1,V_0_danger_val), V_0_danger_val >= 3,
        place(V0), place(V1).
\end{lstlisting}
The bound is \code{3} because that is the smallest score among the streets labelled risky. Any
threshold in \code{(1,3]} separates this data equally well, and FastLAS can only propose bounds at
values that actually occur, so it takes the lowest positive one.

\subsection*{Worked solutions: With pen and paper}

\paragraph{Solution 16.}
Four rules: \code{p.}, \code{p :- a.}, \code{p :- b.} and \code{p :- a, b.} \ Every subset of the
two declared body literals, with \code{p} as the head. ILASP reports nine because it generates the
negated form of each \code{#modeb} for you, so its body literals are \code{a}, \code{b},
\code{not a} and \code{not b}, which yield five further rules
(\code{p :- not a.}, \code{p :- not b.}, \code{p :- a, not b.}, \code{p :- b, not a.} and
\code{p :- not a, not b.}). In FastLAS negation is opt-in, so to get the same nine you would add
\code{#modeb(not a).} and \code{#modeb(not b).}

\paragraph{Solution 17.}
\code{5}. The bias charges 2 for the head and 3 for each body literal, and the learned rule
\code{q :- r.} has one of each, so $2 + 3 = 5$. Both examples are hard and both are covered, so
nothing is added for uncovered examples.

\paragraph{Solution 18.}
The score is \code{1}. With the weights swapped, FastLAS compares two options. Learning \code{p.}
covers \code{e1} but abandons \code{e2}, costing $1$ for the rule plus $3$ for the abandoned
example, so $4$. Learning nothing covers \code{e2}, since with no rule \code{p} never holds, and
abandons \code{e1} for its penalty of $1$; the empty hypothesis itself costs nothing, so the total
is $1$. The second is cheaper, and an empty answer is FastLAS telling you that on this data, at
these prices, no rule is worth its cost.

\subsection*{Worked solutions: At the keyboard}

\paragraph{Solution 19.}
Add \code{#neg(n1, {roadworthy(c2)}, {}, {}).} A negative example demands that \emph{no} answer
set makes \code{c2} roadworthy, which the over-general rule violates, so the exception is forced
into the body:
\begin{lstlisting}[style=result]
roadworthy(V0) :- not failed(V0), car(V0).
\end{lstlisting}
Because the task now contains a \code{#neg}, it must be run with \code{--nopl}.

\paragraph{Solution 20.}
The answer flips at \code{3}. FastLAS is comparing two options: learn \code{p.} and abandon
\code{e2}, which costs $1 + 1 = 2$, or learn nothing and abandon \code{e1}, which costs
whatever penalty \code{e1} carries. The rule wins only once that penalty is strictly greater
than~2. At \code{e1@2} the two options tie at~2 and either answer may come back, on different
runs of the same file; from \code{e1@3} the rule wins outright, with a score of~2.

\paragraph{Solution 21.}
Make the predicate you are learning one that no example mentions. In the task below only
\code{alarm} is ever observed; \code{suppressed} is never in an inclusion or an exclusion, and
reaches the observations only through the background rule.
\begin{lstlisting}[style=fastlas]
% Solvable under --nopl, UNSATISFIABLE under --opl. The target predicate
% "suppressed" appears in no example: only "alarm" is ever observed, and
% suppressed reaches it through the background rule.
alarm :- fault, not suppressed.

#modeh(suppressed).
#modeb(safe).

#pos(e1, {},      {alarm}, { fault. safe. }).
#pos(e2, {alarm}, {},      { fault. }).

#bias("penalty(1, head)    :- in_head(X).").
#bias("penalty(1, body(X)) :- in_body(X).").
\end{lstlisting}
\begin{lstlisting}[style=shell]
$ FastLAS --opl examples/ex31_nopl_only.las
UNSATISFIABLE
$ FastLAS --nopl examples/ex31_nopl_only.las
suppressed :- safe.
\end{lstlisting}
This is exactly the non-observational case of Section~\ref{sec:oplnopl}: \code{--opl} has nothing
to generalise from, because the target never appears in the data it is given, while \code{--nopl}
abduces what \code{suppressed} would have to be. A second, quite different construction also works:
put a \code{#neg} in the task, since those need \code{--nopl} too
(Section~\ref{sec:context}).

\paragraph{Solution 22.}
Ask for a hypothesis that describes a \emph{space} of possibilities rather than one conclusion.
The three-coin task does it: each coin lands heads or tails, and no deterministic set of rules can
produce both observed flips. In ILASP:
\begin{lstlisting}[style=fastlas]
% An ILASP task, not a FastLAS one. Run it with
%   ILASP --version=4 <file>
% Three coins are flipped twice. The only hypothesis covering both flips
% is a pair of rules forming an even loop through negation, which ILASP
% builds and FastLAS does not.
% The FastLAS rendering of the same task,
% examples/solutions/sol16_coins.las, returns UNSATISFIABLE.
coin(c1). coin(c2). coin(c3).

#modeh(heads(var(coin))).
#modeh(tails(var(coin))).
#modeb(heads(var(coin))).
#modeb(tails(var(coin))).
#modeb(coin(var(coin))).
#maxv(1).

#pos({heads(c1), tails(c2), heads(c3)},
     {tails(c1), heads(c2), tails(c3)}).
#pos({heads(c1), heads(c2), tails(c3)},
     {tails(c1), tails(c2), heads(c3)}).
\end{lstlisting}
\begin{lstlisting}[style=shell]
$ ILASP --version=4 examples/ex32_coins_ilasp.las
tails(V1) :- coin(V1); not heads(V1).
heads(V1) :- coin(V1); not tails(V1).
\end{lstlisting}
Those two rules are an \emph{even loop through negation}: each holds exactly when the other fails,
so the program has two answer sets per coin, and the flips are covered by choosing between them.
The same task posed to FastLAS is \code{examples/solutions/sol16_coins.las}, and it answers
\code{UNSATISFIABLE} under both algorithms:
\begin{lstlisting}[style=shell]
$ FastLAS --opl  examples/solutions/sol16_coins.las
UNSATISFIABLE
$ FastLAS --nopl examples/solutions/sol16_coins.las
UNSATISFIABLE
\end{lstlisting}
FastLAS does not construct such loops, and this is the boundary rather than an accident of the
encoding: weak constraints and choice rules are out of reach for the same reason
(Section~\ref{sec:preferences}). If the answer you want is a space of possibilities, the task is
ILASP's.

\paragraph{Solution 23.}
FastLAS answers \code{UNSATISFIABLE}, under \code{--opl} and \code{--nopl} alike, and it does so
even if you keep only the first flip. The task gives every coin the same description: the contexts
are empty, the background says only \code{coin(c1). coin(c2). coin(c3).}, and so nothing
distinguishes one coin from another. Any hypothesis FastLAS builds is therefore a set of rules
that treats all three coins alike, and its answer set makes \code{heads} true for all of them or
for none, which no example accepts. Covering the observation needs a hypothesis with \emph{several}
answer sets, one per way the coins could land, and the pair ILASP finds is exactly that: an even
loop through negation, in which \code{heads(V)} holds when \code{tails(V)} does not and vice versa.
FastLAS does not construct such loops. This is the same boundary as the choice rules and weak
constraints of Section~\ref{sec:preferences}, and it is a good instinct to develop: if the answer
you want is a \emph{space} of possibilities rather than one definite conclusion, the task belongs
to ILASP.

\section*{Acknowledgements and statements}
This material grew out of a series of PhD-level lectures on ILASP and FastLAS given by the author,
and reuses the slides, exercises and lecture notes written for them, in particular those from the
intensive course \emph{Logic Programming and Explainable AI} taught at the IECS Doctoral School,
University of Trento, in 2023.

It is an unofficial guide written from a user, for the users. The author is not affiliated with
ILASP LTD, has never participated in the development of ILASP or FastLAS, and speaks here only as
an external user, teacher and document author.

Artificial intelligence was used to reorganise that material into these notes, and for editing,
proofreading, and running and re-running the examples. All content was subsequently checked by hand
by the author, who is responsible for any error that remains. Some will certainly have escaped both
the machine and the author, and corrections are genuinely welcome: please write to
\texttt{fabioaurelio.dasaro@univr.it} or \texttt{fabio.d'asaro.14@ucl.ac.uk}.

\bibliographystyle{acmtrans}
\bibliography{fastlas-guide}

\label{lastpage}
\end{document}